\documentclass[twocolumn,showpacs,preprintnumbers,amsmath,amssymb]{revtex4}
\usepackage{epsfig}
\usepackage{dcolumn}
\usepackage{bm}

\newcommand{\remove}[1]{}

\def\be{\begin{equation}}
\def\ee{\end{equation}}

\newcommand{\beq}{\begin{equation}}
\newcommand{\eeq}{\end{equation}}
\newcommand{\beqa}{\begin{eqnarray}}
\newcommand{\eeqa}{\end{eqnarray}}

\newcommand{\bea}{\begin{array}}
\newcommand{\ea}{\end{array}}

\newcommand{\lambdabbar}{{\mkern0.75mu\mathchar '26\mkern -9.75mu\lambda}}
\newcommand{\hh}{\hslash}

\begin{document}

\title{Cosmological cancellation of the vacuum energy density}

\author{Philippe Brax}
\affiliation{Institut de Physique Th\'eorique,\\
CEA, CNRS, F-91191 Gif-sur-Yvette, C\'edex, France}
\author{Patrick Valageas}
\affiliation{Institut de Physique Th\'eorique,\\
CEA, CNRS, F-91191 Gif-sur-Yvette, C\'edex, France}
\vspace{.2 cm}

\date{\today}
\vspace{.2 cm}

\begin{abstract}

We propose a simple model that provides a dynamical cancellation mechanism of the
vacuum energy density appearing either in the form of a bare cosmological constant,
quantum fluctuations of  matter fields or the result of phase transitions.
This ``conformal compensator model'' is based on a conformal coupling $A(\varphi)$
between the Einstein and the Jordan frames.
This couples a second scalar field $\lambda$ to the trace
of the matter energy-momentum tensor, including the bare cosmological constant,
and serves as a dynamical Lagrange multiplier. As a result, the scalar  $\lambda$  relaxes
to a value which cancels the contributions from the vacuum energy density to the Friedmann
equation, and adjusts itself  to changes of the vacuum energy density after matter phase
transitions.
This circumvents Weinberg's theorem through the time dependence of the background
scalar field $\varphi$.
The radiation era, where the vacuum energy is annulled, is recovered in a natural manner.
It is also possible to recover the matter era, via  a tracking of the matter energy density by
the scalar field, as well as the inflationary and dark energy eras, which correspond to regimes
where the cancellation mechanism becomes inefficient. This suggests that inflation, dark energy,
and the annulation of the vacuum energy density, could be related to the same mechanism.
In this setting, the usual fine-tuning of the vacuum energy is avoided, although the onset of the
dark energy era at the appropriate time is not explained.

\end{abstract}

 \vskip2pc

\maketitle

\section{Introduction}
\label{sec:Introduction}

The discovery of the acceleration of the expansion of the Universe at low
redshifts \cite{Perlmutter:1998hx,Riess:1998cb} has renewed the interest in the
cosmological constant problem \cite{Weinberg:1988cp}.
This has led to the investigation of many theories that attempt to address the ``new''
cosmological constant problem, i.e., to give rise to a late-time self-acceleration that
matches observational data \cite{Copeland:2006wr,Amendola:2012ys}.
This is typically achieved by adding a new fluid component to the energy budget of the
Universe, the so-called ``quintessence'' \cite{Ratra:1987rm,Wetterich:1987fm}, or modifying
the laws of gravity on large scales \cite{Clifton:2011jh,Koyama:2015vza}.
However, solar-system and astrophysical data (e.g., pulsar binaries \cite{Hulse:1974eb})
strongly constrain modifications of gravity on small scales \cite{Bertotti:2003rm,Williams:2012a}.
This implies nonlinear screening mechanisms
\cite{Vainshtein:1972sx,Khoury:2003aq,Khoury:2003rn,Pietroni:2005pv,Olive:2007aj,Babichev:2009ee,Hinterbichler:2010es,Brax:2010gi}
to ensure convergence to General Relativity in small-scale and high-density environments.
The recent observation of the equality of the speeds of light and of gravitational waves \cite{TheLIGOScientific:2017qsa} has also further restricted the space
of modified-gravity theories \cite{Creminelli:2017sry,Sakstein:2017xjx}.
Besides, observations  of both
the background dynamics and the large-scale structures have not detected any
significant deviation from the standard $\Lambda$-CDM scenario so far
\cite{Ishak2019}.

Most of these models, as well as the standard $\Lambda$-CDM model,
do not address the more fundamental ``old'' cosmological constant problem,
which is the question of why the observed cosmological constant is so small,
as compared with expectations from particle physics and quantum field theory
\cite{Weinberg:1988cp,Nobbenhuis:2004wn,Padilla_2015}.
More precisely, the issue is the sensitivity of the cosmological constant
to radiative corrections. Indeed, loop corrections to the
vacuum energy density typically generate contributions of order $m^4$,
where $m$ is the scale of the particles included in the theory.
As the standard model of particle physics already contains particles up to
the TeV scale, this gives a contribution to the vacuum energy density, i.e.,
to the renormalized cosmological constant, that is at least $10^{60}$ times greater
than the observed value.
To match observations, one would need to fine tune the bare cosmological constant,
or counterterm, to the contributions at all orders from all particles.
This is made even more problematic by the matter phase transitions experienced
by the Universe. As the background temperature drops with the cosmological expansion,
the Universe is expected to go through the electroweak and quantum chromodynamic
phase transitions, where the vacuum energy density jumps by amounts of order
$(100 \,{\rm GeV})^4$ and $(0.2 \,{\rm GeV})^4$. Therefore, even if the cosmological
constant had been adjusted to a low or zero value in the primordial Universe,
the tuning would be spoilt after these transitions.

This suggests the existence of a self-tuning mechanism, which tames this
extreme sensitivity to UV physics of the cosmological constant seen by
gravity.
Short-scale modifications of gravity through supersymmetric
large extra dimensional scenarios have been proposed
\cite{Burgess:2013ara}, as well as string-theory scenarios
\cite{Heckman:2019dsj}.
However, as the observed cosmological constant only plays a practical role in cosmology,
and corresponds to an infrared contribution to the gravitational force,
it is natural to look for large-scale physics or cosmological frameworks.
Thus, in the degravitation proposal \cite{ArkaniHamed:2002fu},
extended sources beyond a length scale $L$ are filtered out and do not contribute
to the gravitational force. This actually implies a strong modification of General Relativity,
as it yields a massive graviton that propagates five instead of two degrees of freedom
\cite{Dvali:2007kt}.
Models involving a superfluid component, associated with Lorentz-violating massive
gravity, have also been proposed \cite{Khoury:2018vdv}.
More conservative modifications of gravity are provided by
scalar-tensor theories, e.g. within the Horndeski class \cite{Horndeski:1974wa}.
A few of such models have been proposed
\cite{Ford:1987de,Dolgov:2008rf,Evnin:2018zeo,Charmousis:2011bf,Copeland:2012qf,Nunes:2017bwb,Appleby:2018yci},
where the scalar field is coupled to various curvature terms, that can lead to a self-tuning
mechanism. Then, the cosmological constant can be screened while solutions
that mimic the various cosmological eras can be associated with different fixed points
of the dynamics.
Modifications of gravity are actually very delicate, because of the
exquisite match between the predictions of General Relativity and measurements
of gravity on small astrophysical scales and the solar system.
This calls for nonlinear screening mechanisms, that can limit the range where one can
derive practical predictions, and can raise issues regarding UV completions
\cite{Adams:2006aa,Bellazzini:2017fep}.
Then, going to even larger scales, one can consider the Universe, or the full spacetime
volume, as a whole.
A radical proposal is to introduce a mirror universe \cite{Linde:1988ws},
with negative energy particles so that the two cosmological constants (given
by an average over all spacetime of the other universe content)
cancel each other.
Alternatively, keeping only one universe, the sequestering mechanism
\cite{Padilla_2014,Kaloper:2014dqa}
introduces global variables. This yields constraint equations that set the bare cosmological
constant to one-fourth of the average over all spacetime of the trace of the
energy-momentum tensor. This automatically cancels the vacuum energy at all loop orders.
This mechanism can also be obtained from a local theory 
\cite{Kaloper:2015jra,Padilla:2018hvp}.
A similar approach, based on global variables, can also relate the observed cosmological
constant to the formation of nonlinear structures at late times
\cite{Lombriser:2018aru,Lombriser:2019jia}.

In this paper, we present a new model that tackles this ``old'' cosmological problem.
In a manner similar to the sequestering mechanism devised by \cite{Padilla_2014},
we use a conformal mapping between the Jordan-frame metric seen by matter
and the Einstein-frame metric seen by General Relativity (i.e., entering the
Einstein-Hilbert action), to couple a new scalar field $\lambda$ to the trace
$T^\mu_\mu$ of the matter energy-momentum tensor.
Then, as $\lambda$ relaxes to $T^\mu_\mu$, the contributions from the vacuum
energy density to the Friedmann equations are canceled by $\lambda$.
As compared with the sequestering mechanism \cite{Padilla_2014},
which involved global variables (i.e., that do not depend on space or time),
in our framework $\lambda({\bf x},t)$ is a dynamical field.
An advantage is that this ``conformal compensator model'' follows the usual causality
pattern. Whereas
the use of global variables leads to an effective cosmological constant that depends
on both the past and future history of the Universe, through averages over all
spacetime (which then needs to be finite), the field $\lambda({\bf x},t)$ dynamically
responds to the current and past history of the system.
However, this raises the problem that the coupling to $T^\mu_\mu$ also generates
a nondesired coupling to the density of nonrelativistic matter.
This is not surprising, as from the value of the energy-momentum tensor
$T^\mu_\nu$ at a given time, it is not possible to separate without any ambiguity
the contributions from the vacuum, or a classical cosmological constant, from
the matter components (which may have intricate equations of state).
This is circumvented in \cite{Padilla_2014} by the fact that the average over
all spacetime is dominated by the late-time low-energy vacuum energy density,
which stands out as the only component that is not diluted by the expansion in the
far future.
This characterization of the vacuum energy is not possible within our dynamical
framework.
In this paper, we suggest that a possible solution is to use this coupling to
matter to make the scalar field contributions track the matter component, while
remaining subdominant.
Unfortunately, we will find that our explicit implementation is not fully satisfactory,
but  could lead to more efficient models.
On the other hand, $\lambda({\bf x},t)$ being dynamical may offer interesting
possibilities. In particular, it could relate together different eras of the
cosmological history. Thus, the current dark-energy era, the primordial inflationary
stage, and the ``old'' cosmological constant problem, could be related, phases of
accelerated expansion naturally appearing as periods where this same cancellation
mechanism associated with $\lambda$ becomes inefficient.

This paper is organized as follows.
We introduce our model in section~\ref{sec:model}, where we also derive the equations
of motion and describe the generic mechanism that cancels the vacuum energy density.
Next, we study the radiation era in section~\ref{sec:Radiation-era}.
We describe how the system naturally responds to the jumps of the matter vacuum
energy density at the electroweak and quantum chromodynamic phase transitions,
and quickly cancels the new vacuum energy density in the Friedmann equations.
We provide an explicit numerical computation for a simple scalar-field Lagrangian,
which recovers a realistic cosmological history.
We proceed to the matter era in section~\ref{sec:Matter-era}.
We discuss the issues raised by the transition from the radiation era to the matter era,
and the tracking solutions that allow us to recover a matter era, driven by the matter
density component.
We also present an explicit numerical computation.
Then, we consider the dark energy era in section~\ref{sec:Dark-energy-era}.
We explain how periods of accelerated expansions can be easily recovered within our
framework, as periods where the cancellation mechanism stops or becomes inefficient.
We again present an explicit numerical computation for illustrative purposes.
Next, in section~\ref{sec:Inflation-era}, we sketch how our framework could also provide
alternative scenarios for an inflationary era in the primordial universe and its transition
to the radiation era. We give a simple numerical illustration.
Finally, in section~\ref{sec:conclusion}, we conclude and discuss the issues that require
further investigations.

\section{Definition of the model}
\label{sec:model}

\subsection{Definition of the total action}
\label{sec:action}

Let us consider the action
\beq
S = S_{\rm EH} + S_{\rm m} + S_{\varphi,\lambda}
\label{eq:S-def}
\eeq
with
\beq
S_{\rm EH} = \int d^4x \sqrt{-g} \frac{M_{\rm Pl}^2}{2} R ,
\label{eq:S-EH}
\eeq
\beq
S_{\rm m} = \int d^4x \sqrt{- \tilde{g}} \tilde{\cal L}_{\rm m}(\psi^{(i)}_{\rm m},\tilde{g}_{\mu\nu}) ,
\label{eq:S-m}
\eeq
and
\beq
S_{\varphi,\lambda} = \int d^4x \sqrt{-g} \left[ {\cal M}^3 A^4(\varphi) \lambda
+ {\cal M}^4 K(\varphi;X,Y,Z) \right]  ,
\label{eq:S-phi-lambda}
\eeq
where we defined the dimensionless kinetic terms (using the Einstein-frame metric
$g_{\mu\nu}$),
\beq
X = - \frac{\partial^\mu\lambda \partial_\mu\lambda}{2{\cal M}^4} , \;\;\;
Y = - \frac{\partial^\mu\varphi \partial_\mu\lambda}{{\cal M}^4} , \;\;\;
Z = - \frac{\partial^\mu\varphi \partial_\mu\varphi}{2{\cal M}^4} .
\label{eq:X-Y-Z-def}
\eeq
The first term $S_{\rm EH}$ is the usual Einstein-Hilbert action of General Relativity,
written in terms of the Einstein-frame metric tensor $g_{\mu\nu}$.
The second term $S_{\rm m}$ is the matter action (associated with all particles,
including photons and dark matter), where $\psi^{(i)}_{\rm m}$ are the various matter
fields and $\tilde{g}_{\mu\nu}$ is the Jordan-frame metric, seen by matter,
which we define by the conformal rescaling
\beq
\tilde{g}_{\mu\nu} = A^2(\varphi) g_{\mu\nu} , \;\;\; A(\varphi) > 0 .
\label{eq:conformal}
\eeq
Here $\varphi(x)$ is an additional scalar field. We also introduced a second scalar field
$\lambda(x)$, and both scalar fields enter the new term $S_{\varphi,\lambda}$.
Here ${\cal M}$ is a mass parameter that we introduce for dimensional purposes.
We shall check that the cancellation mechanism does not depend on the value of ${\cal M}$.
Indeed, for any constant rescaling factor $\alpha$ the change
${\cal M} \to \alpha {\cal M}$ is absorbed by the change
$\lambda \to \alpha^{-3} \lambda$. This also requires appropriate changes to the function $K$.
Therefore, the choice of ${\cal M}$ corresponds for instance
to a choice of normalization for $\lambda$.
For constant $\varphi$ and $\lambda$, neglecting $S_{\varphi,\lambda}$ we recover
General Relativity and the standard model of particle physics, with a rescaling of the Planck mass
seen by matter in the Jordan frame,
\beq
\tilde{M}^2_{\rm Pl} = M^2_{\rm Pl}/A(\varphi)^2 .
\label{eq:M-Pl}
\eeq
The idea leading to the action (\ref{eq:S-def}) is that $\varphi$ plays the role of a Lagrange
multiplier that enforces the cancellation of the vacuum energy arising from the matter sector
by the second field $\lambda$.
This can be expected by noticing that the action $S$ obeys the symmetry
\beq
\tilde{\cal L}_{\rm m} \to \tilde{\cal L}_{\rm m} - \tilde{V}_{\rm vac} , \;\;\;
\lambda \to \lambda + \tilde{V}_{\rm vac}/{\cal M}^3 , \;\;\;
S \to S ,
\label{eq:S-sym}
\eeq
where we used $\sqrt{-\tilde{g}\mathstrut} = A^4 \sqrt{-g}$ and $\tilde{V}_{\rm vac}$ is any constant
shift of the matter-sector vacuum energy.
This cancellation mechanism arises from the first term in the action
$S_{\varphi,\lambda}$.
The second term is introduced to enlarge the space of solutions and behaviors.
To make it independent of the value of the matter vacuum energy, it only depends
on derivatives of $\lambda$.

In fact, because the matter action couples the scalar field $\varphi$ to the trace
$T^\mu_\mu$ of the matter energy-momentum tensor, the Lagrange multiplier $\varphi$
will ensure the cancellation of all matter contributions to the trace $T^\mu_\mu$.
This exactly cancels any constant vacuum energy density, but also partly cancels the
nonrelativistic matter density. In contrast, the radiation energy-momentum tensor is not
canceled at all because its trace vanishes.
This means that the cancellation mechanism associated with the action (\ref{eq:S-def})
is satisfactory during the radiation era, but can raise problems during the
late matter era and must stop during the dark energy and inflation eras.

\subsection{Equations of motion}
\label{sec:motion}

For simplicity, we consider conformal rescalings such that
$A(\varphi)$ is constant at late times
and it is sufficient to analyze the Einstein equations in the Einstein frame.
Defining the matter energy-momentum tensors in the Einstein and Jordan frames as
\beq
T_{({\rm m}) \mu\nu} = \frac{-2}{\sqrt{-g}} \frac{\delta S_{\rm m}}{\delta g^{\mu\nu}} , \;\;\;
\tilde{T}_{({\rm m}) \mu\nu} = \frac{-2}{\sqrt{-\tilde{g}}} \frac{\delta S_{\rm m}}{\delta \tilde{g}^{\mu\nu}} ,
\label{eq:T-T_tilde}
\eeq
we have $T^{\mu}_{({\rm m})\nu} = A^4 \, \tilde{T}^{\mu}_{({\rm m})\nu}$
and the Einstein equations write in the Einstein frame as
\beq
M^2_{\rm Pl} G^{\mu}_{\nu} = A^4 \tilde{T}^{\mu}_{({\rm m}) \nu} + T^{\mu}_{(\varphi,\lambda) \nu} ,
\eeq
where $T^{\mu}_{(\varphi,\lambda) \nu}$ is the energy-momentum tensor associated with the
scalar-fields action $S_{\varphi,\lambda}$.
We write the matter energy-momentum tensor in the Jordan frame as the sum of three
components, the vacuum energy density $\tilde{V}_{\rm vac}$, the nonrelativistic matter
density $\tilde{\rho}$, with negligible pressure, and the radiation density and pressure,
$\tilde{\rho}_{\gamma}$ and $\tilde{p}_{\gamma}=\tilde{\rho}_{\gamma}/3$.
We include a possible cosmological constant into the definition of the vacuum
energy density.
This gives
\beq
\tilde{T}^0_{({\rm m})0} =  - \tilde{V}_{\rm vac} - \tilde{\rho} - \tilde{\rho}_{\gamma} , \;\;\;
\tilde{T}^i_{({\rm m})i} =   - \tilde{V}_{\rm vac} + \tilde{\rho}_{\gamma} / 3 ,
\eeq
while nondiagonal elements vanish.
By definition, between phase transitions the vacuum energy density is constant while the
nonrelativistic and relativistic densities decrease as
\beq
\tilde\rho = \frac{\tilde\rho_0}{\tilde{a}^3} , \;\;\;
\tilde\rho_\gamma = \frac{\tilde\rho_{\gamma 0}}{\tilde{a}^4} , \;\;\;
\tilde{a} = A a ,
\eeq
where $\tilde{a}$ is the Jordan-frame scale factor.
At a phase transition, which can lead to a jump of the vacuum energy density,
the values of $\tilde\rho_0$ and $\tilde\rho_{\gamma 0}$ may also jump
as some energy can be exchanged between the vacuum energy and the matter components.
However, in this section, we focus on the behavior in between phase transitions, where
the vacuum energy density $\tilde{V}_{\rm vac}$ is constant.
The scalar-field energy-momentum tensor reads
\beqa
&& T^{\mu}_{(\varphi,\lambda)\nu} = \left[ {\cal M}^3 A^4 \lambda + {\cal M}^4 K \right]
\delta^{\mu}_{\nu} + \frac{\partial K}{\partial X} \partial^\mu\lambda \partial_\nu\lambda
\nonumber \\
&& + \frac{\partial K}{\partial Y} ( \partial^{\mu}\varphi \partial_{\nu}\lambda
+ \partial^{\mu}\lambda \partial_{\nu}\varphi )
+ \frac{\partial K}{\partial Z} \partial^\mu\varphi \partial_\nu\varphi .
\label{eq:T-varphi-lambda}
\eeqa
Therefore, the Einstein equations read
\beqa
3 M^2_{\rm Pl} {\cal H}^2 & = & a^2 A^4 ( \tilde{V}_{\rm vac} + \tilde\rho
+ \tilde\rho_{\gamma} - {\cal M}^3 \lambda ) - a^2 {\cal M}^4 K \nonumber \\
&& + \frac{\partial K}{\partial X} \lambda'^2
+ 2 \frac{\partial K}{\partial Y} \lambda' \varphi'
+ \frac{\partial K}{\partial Z} \varphi'^2
\label{eq:EE00}
\eeqa
and
\beq
M^2_{\rm Pl} ( {\cal H}^2 + 2 {\cal H}' ) = a^2 A^4 ( \tilde{V}_{\rm vac}
- \tilde\rho_{\gamma}/3 - {\cal M}^3 \lambda ) - a^2 {\cal M}^4 K ,
\label{eq:EEii}
\eeq
where the primes denote the derivative with respect to the conformal time $\tau$ and
${\cal H} = d\ln a/d\tau$ is the conformal Hubble expansion rate.

The derivatives of the action with respect to the scalar fields $\varphi$
and $\lambda$ give the equations of motion
\beqa
&& {\cal M}^4 \frac{\partial K}{\partial\varphi} - a^{-4} \partial_{\tau} \left[
a^2 \left( \frac{\partial K}{\partial Y} \lambda'
+ \frac{\partial K}{\partial Z} \varphi' \right) \right] =
 \nonumber \\
&& 4 A^3 \frac{dA}{d\varphi}
(\tilde{V}_{\rm vac}+\tilde\rho/4 - {\cal M}^3 \lambda)
\label{eq:dS-dphi}
\eeqa
and
\beq
a^{-4} \partial_\tau \left[ a^2 \left( \frac{\partial K}{\partial X} \lambda'
+ \frac{\partial K}{\partial Y} \varphi' \right) \right] = {\cal M}^3 A^4 ,
\label{eq:dS-dlambda}
\eeq
while the kinetic factors are
\beq
X = \frac{\lambda'^2}{2{\cal M}^4 a^2} , \;\;\;
Y = \frac{\varphi'\lambda'}{{\cal M}^4 a^2} , \;\;\;
Z = \frac{\varphi'^2}{2{\cal M}^4 a^2} .
\eeq

\subsection{Cancellation mechanism in the radiation era}
\label{sec:cancellation-radiation}

We can see at once in Eqs.(\ref{eq:EE00})-(\ref{eq:dS-dlambda}) the cancellation
of the vacuum energy density in the radiation era.
Thus, let us neglect the nonrelativistic matter density $\tilde\rho$
and consider a constant vacuum energy density
$\tilde{V}_{\rm vac}$,
\beq
\tilde\rho = 0 , \;\;\;  \tilde{V}_{\rm vac} = {\rm constant} .
\eeq
Then, the equation of motion (\ref{eq:dS-dphi}) has the constant solution
\beq
\lambda = \tilde{V}_{\rm vac}/{\cal M}^3 , \;\;\; \lambda'=0 ,
\label{eq:lambda-Vvac}
\eeq
provided the kinetic function $K$ satisfies
\beq
\frac{\partial K}{\partial\varphi} = 0 \;\; \mbox{and} \;\;
\frac{\partial K}{\partial Z} = 0 \;\; \mbox{when} \;\; \lambda' = 0 .
\label{eq:K=0-1}
\eeq
Then, the Einstein equations (\ref{eq:EE00}) and (\ref{eq:EEii}) become
\beq
3 M^2_{\rm Pl} {\cal H}^2 = a^2 \rho_{\gamma} , \;\;\;
M^2_{\rm Pl} ( {\cal H}^2+2{\cal H}') = - a^2 \rho_{\gamma}/3 ,
\label{eq:Friedmann}
\eeq
where the Einstein-frame radiation density is
\beq
\rho_{\gamma} = A^4 \tilde\rho_\gamma = \tilde\rho_{\gamma 0} /a^4 ,
\eeq
provided the kinetic function $K$ also satisfies
\beq
K = 0 \;\; \mbox{when} \;\; \lambda' = 0 .
\label{eq:K=0-2}
\eeq
Then, we recover the standard Friedmann equations of the radiation era.

The vacuum energy $\tilde{V}_{\rm vac}$ has been canceled by ${\cal M}^3\lambda$,
as the scalar field $\varphi$ acts as a Lagrange multiplier that enforces the
constraint (\ref{eq:lambda-Vvac}).
The equation of motion (\ref{eq:dS-dlambda}) provides the evolution of $\varphi$.
At this stage, the role of the kinetic term in the scalar field action
(\ref{eq:S-phi-lambda}) is only to make sure that the equation of motion
(\ref{eq:dS-dlambda}) does not imply $A=0$, as would be the case if it were absent.
It does not spoil the vacuum energy cancellation, as long as the latter is constant,
if its effect vanishes for constant $\lambda$ following the conditions
(\ref{eq:K=0-1}) and (\ref{eq:K=0-2}).
The cancellation works for any value of the vacuum energy density
$\tilde{V}_{\rm vac}$ and does not depend on the value of the mass parameter
${\cal M}$, which disappears from the Friedmann equations.

The manner by which this scenario evades the well-known no-go theorem by Weinberg
\cite{Weinberg:1988cp}
can be seen from Eq.(\ref{eq:dS-dlambda}). If we look for static solutions in the
Minkowski background, so that time derivatives vanish, Eq.(\ref{eq:dS-dlambda})
implies at once $A=0$, hence, the matter action vanishes.
This corresponds to Weinberg's result.
In our case, we avoid a vanishing $A$ thanks to the nonzero time derivatives on the
left-hand side. This is because we solve the cosmological constant problem within
a cosmological setting, which implies nonzero time derivatives of the scale factor $a$
as the Universe is expanding. Moreover, the background field $\varphi$ also
evolves with time. Note that in a cosmological framework, because the Universe is not
static there is no reason to require static background fields. In this respect,
our solution of the cosmological constant problem is related to the cosmological
framework of our Universe.
In particular, the Minkowski limit, which applies to laboratory experiments,
is understood as the limit of the FLRW metric over short time scales and small lengths.
But the resolution of the cosmological constant problem must be taken into account
in the exact FLRW metric, before taking the local Minkowski limit.
This way out of Weinberg's no-go theorem is shared by other self-tuning models
\cite{Charmousis:2011bf,Copeland:2012qf,Nunes:2017bwb,Appleby:2018yci},
which also require time-dependent background fields.
An alternative is to introduce a spatial dependence for some background fields
\cite{Khoury:2018vdv}, or Lorentz-violating theories.

\subsection{Dimensionless variables}
\label{sec:dimensionless}

It is convenient to write the equations of motion in terms of dimensionless variables.
Thus, we define the dimensionless density parameters and the reduced
Hubble expansion rate
\beqa
&& \frac{\tilde{V}_{\rm vac}}{3 M_{\rm Pl}^2 H_0^2} = \Omega_{\rm vac 0} , \;\;\;
\frac{\tilde\rho}{3 M_{\rm Pl}^2 H_0^2} = \frac{\Omega_0}{A^3 a^3} , \nonumber \\
&& \frac{\tilde\rho_\gamma}{3 M_{\rm Pl}^2 H_0^2} = \frac{\Omega_{\gamma 0}}{A^4 a^4} , \;\;\;
\hh = \frac{H}{H_0} .
\eeqa
The parameters $\Omega_{i 0}$ are constant during most of the history of the Universe,
but can vary during phase transitions.
We also define the dimensionless scalar fields
\beq
\hat{\varphi} = \frac{\varphi}{M_{\rm Pl}} , \;\;\;
\hat{\lambda} = \frac{{\cal M}^3 \lambda}{3 M_{\rm Pl}^2 H_0^2} ,
\eeq
the rescaled kinetic factors,
\beq
\hat{X} = \frac{\hh^2}{2} \left( \frac{d\hat\lambda}{d\eta} \right)^2 ,\;\;\;
\hat{Y} = \hh^2 \frac{d\hat\lambda}{d\eta} \frac{d\hat\varphi}{d\eta}  , \;\;\;
\hat{Z} = \frac{\hh^2}{2} \left( \frac{d\hat\varphi}{d\eta} \right)^2 ,
\label{eq:XYZ-hat}
\eeq
and the rescaled kinetic function
\beq
\hat{K}(\hat\varphi;\hat{X},\hat{Y},\hat{Z}) = \frac{{\cal M}^4}{3 M_{\rm Pl}^2 H_0^2}
K(\varphi;X,Y,Z) .
\eeq
Using the dimensionless time coordinate $\eta = \ln(a)$, the Einstein equations
(\ref{eq:EE00})-(\ref{eq:EEii}) give
\beqa
\hh^2 & = & A^4 ( \Omega_{\rm vac 0} - \hat\lambda ) + \frac{A\Omega_0}{a^3}
+ \frac{\Omega_{\gamma 0}}{a^4} - \hat{K}  \nonumber \\
&& + 2 \hat{X} \frac{\partial \hat{K}}{\partial \hat{X}}
+ 2 \hat{Y} \frac{\partial \hat{K}}{\partial \hat{Y}}
+ 2 \hat{Z} \frac{\partial \hat{K}}{\partial \hat{Z}}
\hspace{1cm}
\label{eq:EE00-hat}
\eeqa
and
\beqa
2 \hh^2 \frac{d\ln \hh}{d\eta} & = & - 3 \frac{A\Omega_0}{a^3}
- 4 \frac{\Omega_{\gamma 0}}{a^4} - 6 \hat{X} \frac{\partial \hat{K}}{\partial \hat{X}}
 \nonumber \\
&& - 6 \hat{Y} \frac{\partial \hat{K}}{\partial \hat{Y}}
- 6 \hat{Z} \frac{\partial \hat{K}}{\partial \hat{Z}}  ,
\hspace{0.8cm}
\label{eq:EEii-hat}
\eeqa
while the scalar-field equations (\ref{eq:dS-dphi})-(\ref{eq:dS-dlambda}) read as
\beqa
&& \frac{\partial \hat{K}}{\partial\hat\varphi} - a^{-3} \hh \frac{d}{d\eta} \left[
a^3 \hh \left( \frac{\partial \hat{K}}{\partial \hat{Y}} \frac{d\hat\lambda}{d\eta}
+ \frac{\partial \hat{K}}{\partial \hat{Z}} \frac{d\hat\varphi}{d\eta} \right) \right] =
 \nonumber \\
&& 4 A^3 \frac{dA}{d\hat\varphi}
\left( \Omega_{\rm vac0} - \hat\lambda +\frac{\Omega_0}{4 A^3 a^3} \right)
\label{eq:dS-dphi-hat}
\eeqa
and
\beq
a^{-3} \hh \frac{d}{d\eta} \left[ a^3 \hh \left( \frac{\partial \hat{K}}{\partial \hat{X}} \frac{d\hat\lambda}{d\eta}
+ \frac{\partial \hat{K}}{\partial \hat{Y}} \frac{d\hat\varphi}{d\eta} \right) \right] = A^4 .
\label{eq:dS-dlambda-hat}
\eeq
In the following, we work with these dimensionless quantities and omit the hats to simplify notations.

\subsection{Exponential conformal coupling and power-law kinetic function}
\label{sec:exponential}

For simplicity, in this paper we only consider exponentials and power laws for the
conformal coupling function $A$ and the kinetic function $K$.
More precisely, we take a simple exponential for $A(\varphi)$,
\beq
A(\varphi) = A_\star e^{\nu_A \varphi} , \;\;\; A_\star > 0 ,
\label{eq:A-rad-def}
\eeq
while for $K(\varphi;X,Y,Z)$ we take the separable form
\beq
K(\varphi;X,Y,Z) = K_X e^{\nu_X\varphi} X^{\gamma} + K_Y Y ,
\label{eq:K-general}
\eeq
with $\gamma > 0$.
We shall take the parameters $A_\star, \nu_i, K_i, \gamma$ constant for most of the expansion history
of the Universe, but allow them to vary between different eras.
In more complex scenarios, they would only be effective coefficients that provide approximations
of the kinetic function over limited ranges, and smoothly vary with the arguments $\varphi,X,Y$
and $Z$.
For simplicity, we do not include a component of the form $K_Z Z$, because we can already
recover interesting cosmological behaviors in the subclass $K_Z=0$.
It appears that the kinetic functions (\ref{eq:K-general}) are the simplest choice that can
reproduce all cosmological eras, from the inflationary stage to the current dark-energy era.

As the kinetic function $K$ does not depend on $\lambda$, the equations of motion only depend
on the difference $\lambdabbar$ between $\lambda$ and the vacuum energy density,
\beq
\lambdabbar = \lambda - \Omega_{\rm vac0} ,
\label{eq:lambda-bar-def}
\eeq
as can be checked in Eqs.(\ref{eq:EE00-hat})-(\ref{eq:dS-dlambda-hat}).
This is the property that ensures the cancellation of the vacuum energy density, independently
of its value.
Except at matter phase transitions, we shall take the matter vacuum energy density
$\Omega_{\rm vac0}$ to be constant. Then, it will be convenient to write the equations
of motion in terms of $\lambdabbar$, and most of the discussions below will use
$\lambdabbar$.

\section{Early radiation era}
\label{sec:Radiation-era}

\subsection{Equations of motion}
\label{sec:motion-rad}

We now consider in more detail the radiation era and the cancellation mechanism of the
vacuum energy density.
In particular, to check its efficiency we must go beyond the constant-$\lambda$
solution (\ref{eq:lambda-Vvac}) and verify that perturbations decay.
To simplify the analysis, we take $K_Y=0$ in the general class (\ref{eq:K-general}),
and we focus on the simpler kinetic functions
\beq
K(\varphi;X,Y,Z) = K_X e^{\nu_X \varphi} X^{\gamma} , \;\;\; \gamma > 0 ,
\label{eq:K-rad}
\eeq
which satisfy the constraints (\ref{eq:K=0-1}) and (\ref{eq:K=0-2}).
They do not depend on $Y$ and $Z$ and the dependence on $\varphi$ and $X$ factorizes.
From Eq.(\ref{eq:XYZ-hat}) we have $X \geq 0$.
Then, the equations of motion of the scalar fields simplify as
\beq
\nu_X K_X e^{\nu_X\varphi} X^{\gamma} = - 4 \nu_A A_\star^4 e^{4\nu_A\varphi} \lambdabbar
+ \nu_A A_\star e^{\nu_A\varphi} \frac{\Omega_0}{a^3}  ,
\label{eq:dS-dphi-rad}
\eeq
and
\beqa
&& \gamma K_X e^{\nu_X\varphi} X^{\gamma-1} \hh^2 \left[
\left( 3 + (2\gamma-1) \frac{d\ln\hh}{d\eta} + \nu_X \frac{d\varphi}{d\eta} \right) \frac{d\lambdabbar}{d\eta}
\right. \nonumber \\
&& \left. + (2\gamma-1) \frac{d^2\lambdabbar}{d\eta^2} \right] = A_\star^4 e^{4\nu_A\varphi} .
\label{eq:dS-dlambda-rad}
\eeqa
Thus, Eq.(\ref{eq:dS-dphi-rad}) becomes a constraint equation for $\varphi$,
as there is no kinetic term over $\varphi$, while Eq.(\ref{eq:dS-dlambda-rad}) is a nonlinear
second-order equation of motion for $\lambdabbar$.

\subsection{Relaxation solution}
\label{sec:relaxation}

We will perform an exact numerical computation in section~\ref{sec:radiation-numerical-1} below,
but in this section we present an analytic study of the solutions that appear in the radiation era.
We derive explicit solutions and check their linear stability. This allows us to obtain the
range of the parameters $\nu_X$ and $\gamma$ of the kinetic function (\ref{eq:K-rad})
that give rise to the required scalings and stability conditions.
For this purpose, we can neglect the nonrelativistic matter density.
In terms of the effective matter density parameter, this means $\Omega_0=0$.
We have seen in section~\ref{sec:cancellation-radiation} that for a constant vacuum energy density,
the constant solution (\ref{eq:lambda-Vvac}) provides a radiationlike expansion for the
Einstein-frame scale factor. Since we aim at building solutions where the scalar fields
are subdominant in the Friedmann equations (except temporarily at phase transitions),
we take the Hubble expansion rate to follow the radiation era scaling,
\beq
\hh = h_\star e^{-2\eta} ,
\label{eq:h-2eta-rad}
\eeq
where $h_\star$ is an irrelevant proportionality factor.

With $\Omega_0=0$, the constraint equation (\ref {eq:dS-dphi-rad}) gives for $\varphi$
the explicit expression
\beq
\varphi = \frac{1}{4\nu_A-\nu_X} \ln \left[ \frac{-\sigma K_X}{A_\star^4 \lambdabbar}
\left( \frac{h_\star^2}{2} e^{-4\eta} \left( \frac{d\lambdabbar}{d\eta} \right)^2 \right)^\gamma \; \right] ,
\label{eq:phi-rad}
\eeq
while Eq.(\ref{eq:dS-dlambda-rad}) leads to
\beq
\frac{d^2\lambdabbar}{d\eta^2} +  \frac{5-5\sigma-4\gamma}{\sigma+2\gamma-1}
\frac{d\lambdabbar}{d\eta} - \frac{\sigma}{2\gamma} \lambdabbar^{-1}
\left( \frac{d\lambdabbar}{d\eta} \right)^2 = 0 ,
\label{eq:d2lambda-rad}
\eeq
where we introduced the ratio
\beq
\sigma \equiv \frac{\nu_X}{4\nu_A} .
\label{eq:sigma-def}
\eeq
The choice (\ref{eq:K-rad}) implies that $\lambdabbar$ and $d\lambdabbar/d\eta$ are
nonzero, that is, the scalar field $\lambda$
has not completely relaxed to the solution (\ref{eq:lambda-Vvac}).
These equations of motion are nonlinear, but we can look for a simple solution of the form
\beq
\varphi = \varphi_\star+\mu_\varphi\eta , \;\;\;
\lambdabbar = \lambda_\star e^{\mu_\lambda\eta} .
\label{eq:scalar-rad}
\eeq
Substituting into Eqs.(\ref{eq:phi-rad})-(\ref{eq:d2lambda-rad}), we obtain the three
constraints
\beqa
&& \varphi_\star = \frac{1}{4\nu_A-\nu_X} \ln \left[ \frac{-\sigma K_X}{A_\star^4}
\left( \frac{h_\star^2 \mu_\lambda^2}{2} \right)^\gamma \lambda_\star^{2\gamma-1} \right] ,
\nonumber \\
&& \mu_\lambda = \frac{2\gamma (5-5\sigma-4\gamma)}
{(\sigma-2\gamma) (\sigma+2\gamma-1)} , \nonumber \\
&& \mu_\varphi = \frac{\gamma (2\sigma+10\gamma-5)}{2\nu_A(\sigma-2\gamma) (\sigma+2\gamma-1)} .
\label{eq:rad-contraint-3}
\eeqa
This determines the two coefficients $\mu_i$ and sets the normalization of $\lambdabbar$ in terms
of the normalization of $\varphi$.
Indeed, as the nonlinear equation of motion (\ref{eq:d2lambda-rad}) is actually homogeneous of
degree one, any constant rescaling of $\lambdabbar$ provides a new solution.

For the cancellation of the vacuum energy density to occur, we require that the deviation of
$\lambda$ from $\Omega_{\rm vac0}$ decays, that is,
\beq
\mu_\lambda < 0 , \;\; \mbox{hence} \;\; \frac{5-5\sigma-4\gamma}{(\sigma-2\gamma) (\sigma+2\gamma-1)}
< 0 .
\label{eq:mu-lambda-negative}
\eeq
To ensure that the Friedmann equation (\ref{eq:EE00-hat}) remains close to the radiationlike
behavior (\ref{eq:Friedmann}), we also require the stronger condition that $\lambda-\Omega_{\rm vac0}$
decay faster than $\tilde\rho_{\gamma} \propto A^{-4} a^{-4}$ and that $K$ decay faster than
$\rho_{\gamma} \propto a^{-4}$. Both conditions give the same constraint,
\beq
\frac{2\sigma+\gamma-2}{\sigma+2\gamma-1} < 0 .
\label{eq:mu-lambda-rho-rad}
\eeq
Indeed, for the kinetic and coupling functions (\ref{eq:K-rad}) and (\ref{eq:A-rad-def}),
the equation of motion (\ref{eq:dS-dphi-rad}) reads as
\beq
\sigma K = - A^4 \lambdabbar + A \frac{\Omega_0}{4 a^3} .
\label{eq:K-A-lambda}
\eeq
Then, the conditions $|\lambdabbar| \ll \Omega_{\gamma 0}/(A^4a^4)$
and $|K| \ll \Omega_{\gamma 0}/a^4$ are equivalent, as we take $\Omega_0 = 0$ in the
early radiation era.
We also require that the Jordan-frame scale factor $\tilde{a}=A a$ grows with time.
We obtain
\beq
\tilde{a} \propto a^{\alpha} \;\;\; \mbox{with} \;\;\;
\alpha = \frac{2\sigma^2+2 (\gamma-1) \sigma + 2 \gamma^2- \gamma}{2(\sigma-2\gamma)(\sigma+2\gamma-1)} > 0 .
\label{eq:a-J-a-E-rad}
\eeq

To ensure that the solution (\ref{eq:scalar-rad}) is relevant, we also require that it is a local attractor,
i.e., that it is stable. Thus, we consider the small deviations $\delta\varphi$ and $\delta\lambdabbar$,
\beq
\varphi = \varphi_\star+\mu_\varphi\eta + \delta\varphi , \;\;\;
\lambdabbar = \lambda_\star e^{\mu_\lambda\eta} (1+\delta\lambdabbar) .
\label{eq:scalar-rad-delta}
\eeq
Substituting into the equations of motion (\ref{eq:phi-rad})-(\ref{eq:d2lambda-rad}), we obtain
at linear order
\beq
\frac{d^2\delta\lambdabbar}{d\eta^2}
- \frac{\mu_\lambda (\sigma-2\gamma)}{2\gamma} \frac{d\delta\lambdabbar}{d\eta} = 0 .
\eeq
This gives a constant mode, associated with a change of the normalizations $\varphi_\star$ and
$\lambda_\star$ of the solution (\ref{eq:scalar-rad}), and an exponential mode that decays with respect
to the solution (\ref{eq:scalar-rad}) when
\beq
\sigma > 2 \gamma .
\label{eq:mu-lambda-stable}
\eeq

The combination of the four constraints (\ref{eq:mu-lambda-negative}), (\ref{eq:mu-lambda-rho-rad}),
(\ref{eq:a-J-a-E-rad}) and (\ref{eq:mu-lambda-stable}) gives the two allowed regimes
\beqa
0 < \gamma \leq \frac{5}{14} : && \frac{1-\gamma+\sqrt{1-3\gamma^2}}{2} < \sigma < 1 - \frac{\gamma}{2} ,
\;\;\;  \label{eq:gamma-rad-1}
\\
\frac{5}{14} \leq \gamma < \frac{2}{5} : && 2\gamma < \sigma < 1 - \frac{\gamma}{2} .
\label{eq:gamma-rad-2}
\eeqa
Thus, only the ratio $\sigma=\nu_X/4\nu_A$ and the exponent $\gamma$ are constrained
by these stability requirements. A change of $\nu_A$ at fixed $\sigma$ simply gives
a rescaling of the evolution of the scalar field $\varphi$.
The scalar field $\varphi$ and the conformal factor $A$ between the Einstein and Jordan frames
are constant if $\mu_\varphi=0$. This corresponds to
\beq
\mu_\varphi=0  \;\; \mbox{when} \;\; \sigma = \frac{5}{2} - 5 \gamma , \;\;\;
\frac{1}{3} < \gamma < \frac{5}{14} .
\label{eq:mu-phi=0}
\eeq
As $\lambdabbar$ will typically jump upward at phase transitions, we take $\lambda_\star > 0$.
Then, Eq.(\ref{eq:phi-rad}) implies $K_X<0$.

\subsection{Matter phase transitions}
\label{sec:phase}

We have described in the previous section the smooth evolution of the scalar fields
at constant vacuum energy density.
However, during the radiation era, the Universe is expected to go through several phase
transitions (PT), such as the quantum chromodynamics (QCD) PT at
$T_{\rm QCD} \sim 200 \, {\rm MeV}$, the  electroweak (EW) PT at
$T_{\rm EW} \sim 100 \, {\rm GeV}$, and possibly the grand unification (GUT) PT
at $T_{\rm GUT} \sim 10^{15} \, {\rm GeV}$.
At each transition, we expect the vacuum energy density $\tilde{V}_{\rm vac}$
to jump downward by an amount of order $T^4 \sim \tilde\rho_\gamma$.

In this paper, we are not interested in the details of the phase transitions and
we consider the simpler case of instantaneous and homogeneous transitions.
Then, the vacuum energy density parameter $\Omega_{\rm vac 0}$ jumps from
$\Omega_{\rm vac 1}$ to $\Omega_{\rm vac 2}$ at the transition time $\eta$,
by the amount
\beq
\Delta \Omega_{\rm vac} = \Omega_{\rm vac 2} - \Omega_{\rm vac 1} = - \alpha_{\rm pt}
\frac{\Omega_{\gamma 1}}{A_1^4 a^4} .
\label{eq:Delta-Omega-vac-def}
\eeq
This corresponds to the change of vacuum energy density
$\Delta \tilde{V}_{\rm vac} = - \alpha_{\rm pt} \tilde\rho_{\gamma}$, in the matter Jordan frame,
and we expect $\alpha_{\rm pt} \lesssim 1$.
This leads to a jump of the radiation energy density parameter $\Omega_{\gamma}$,
of the scalar field $\varphi$, and of the first derivative $d\lambda/d\eta$.
Indeed, the Einstein equations are of second order in the metric, and the second Friedmann
equation (\ref{eq:EEii-hat}) enforces the Hubble expansion rate $\hh$ to be continuous.
Next, from the equation of motion (\ref{eq:dS-dlambda-hat}), the scalar field $\lambda$ and
the product $\frac{\partial K}{\partial X} \frac{d\lambda}{d\eta}$
are also continuous. On the other hand, because there is no kinetic term in $\varphi$
for the class of kinetic functions (\ref{eq:K-rad}), the scalar field $\varphi$ is discontinuous
and follows the constraint equation (\ref{eq:dS-dphi-hat}).
This gives the junction conditions at the transition,
\beq
\left( \frac{d\lambda}{d\eta} \right)_2 = \left( \frac{d\lambda}{d\eta} \right)_1
\left( \frac{\lambda - \Omega_{\rm vac2}}{\lambda-\Omega_{\rm vac1}} \right)^{\sigma/(\sigma+2\gamma-1)} ,
\label{eq:dlambda-pt}
\eeq
and
\beqa
\Omega_{\gamma2} - \Omega_{\gamma1} & = & a^4 A_1^4 (\Omega_{\rm vac1}-\lambda)
- a^4 A_2^4 (\Omega_{\rm vac2}-\lambda) \nonumber \\
&& + (2\gamma-1) a^4 (K_1-K_2) .
\label{eq:Omegarad-pt}
\eeqa
The drop of the vacuum energy density is not identically transferred to the radiation energy density
because the scalar field kinetic and coupling functions are also discontinuous at the transition
and enter the energy balance.
We could make the scalar field contributions continuous by including kinetic terms in $(\partial\varphi)^2$
in the kinetic function, but for simplicity we keep the same kinetic function (\ref{eq:K-rad}) throughout
the radiation era.
At the phase transition, the difference $\lambdabbar=\lambda-\Omega_{\rm vac0}$
shows a positive jump, because of the discontinuity of $\Omega_{\rm vac0}$, with
\beq
\lambdabbar_2 = \lambdabbar_1 - \Delta\Omega_{\rm vac} .
\label{eq:lambdabbar-pt}
\eeq
After the phase transition, provided the system remains in the basin of attraction of the solution
(\ref{eq:scalar-rad}), the difference $\lambdabbar$ again decays and $\lambda$ cancels the new
vacuum energy density $\Omega_{\rm vac2}$.

\subsection{Numerical computation}
\label{sec:radiation-numerical-1}

For a numerical computation of the evolution of the fields in the radiation era,
we do not use the approximation (\ref{eq:h-2eta-rad}) as we keep the exact Hubble expansion
rate given by the Friedmann equation (\ref{eq:EE00-hat}).
We also take into account the nonrelativistic matter density $\Omega_0$ in the equation of motion
(\ref{eq:dS-dphi-rad}).

We consider two phase transitions during the radiation era, the electroweak and QCD phase transitions,
which we set at
\beq
T_{\rm EW} = 100 \, {\rm GeV} \;\;\; \mbox{and} \;\;\; T_{\rm QCD} = 200 \, {\rm MeV} .
\eeq
We model them as instantaneous, with a sudden jump of the vacuum energy density
as in Eq.(\ref{eq:Delta-Omega-vac-def}), with
\beq
\alpha_{\rm EW} = 0.1 \;\;\; \mbox{and} \;\;\; \alpha_{\rm QCD} = 0.1 .
\eeq
This is a simplified and somewhat arbitrary choice, but our goal here is simply to check that
the system can handle such phase transitions and restore the cancellation of the vacuum energy density.
We can expect that if this is the case, it would also accommodate more realistic and smoother
phase transitions.

\begin{figure}
\begin{center}
\epsfxsize=8. cm \epsfysize=6 cm {\epsfbox{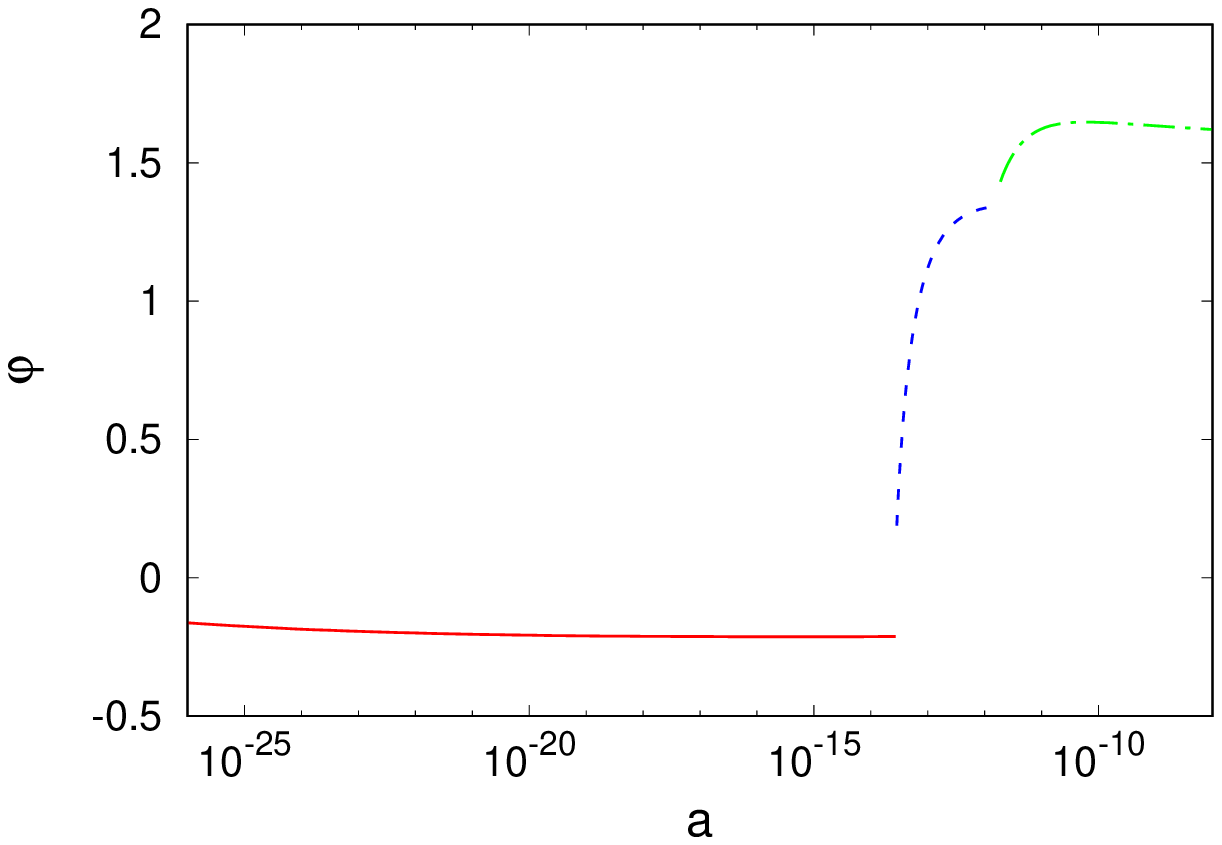}}
\epsfxsize=8. cm \epsfysize=6 cm {\epsfbox{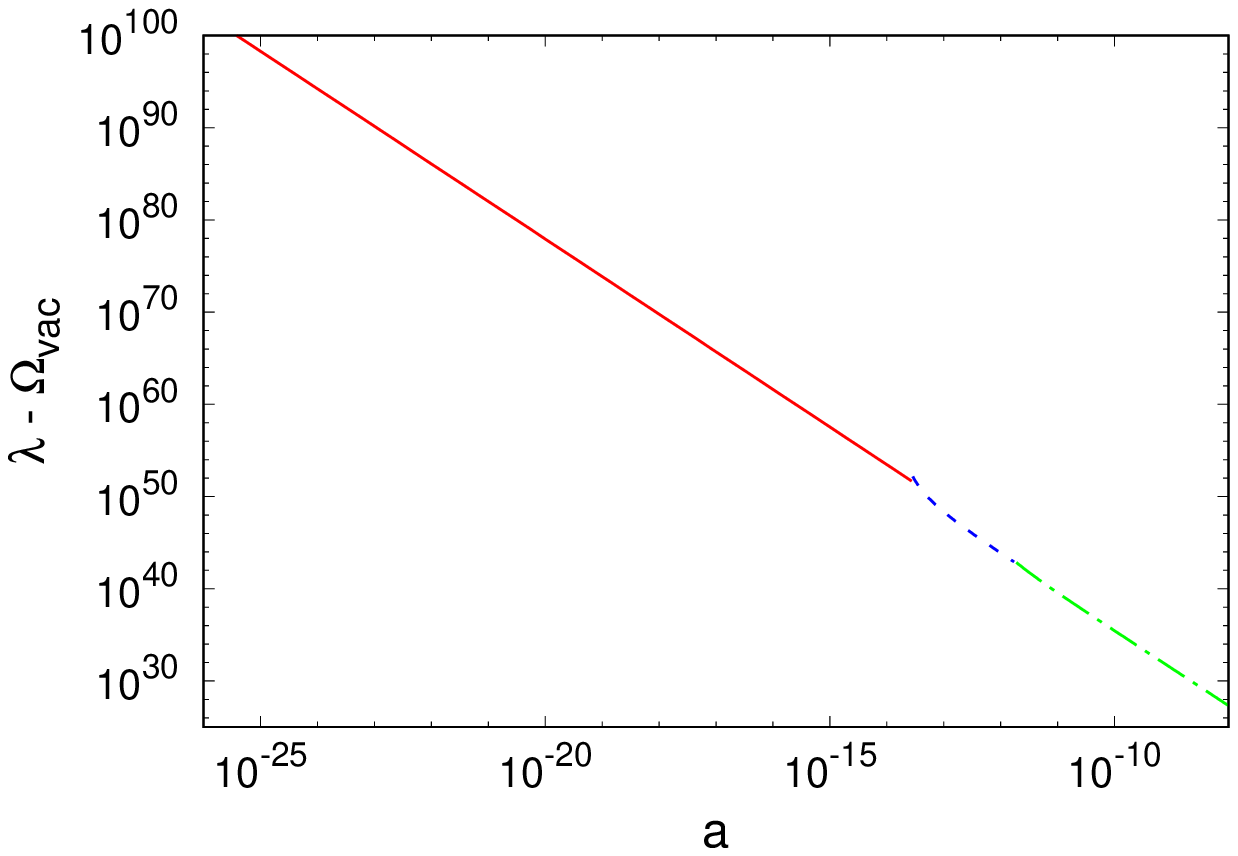}}

\end{center}
\caption{
{\it Upper panel:} scalar field $\varphi$ during the radiation era, as a function
of the scale factor $a$.
The red solid line corresponds to $T>T_{\rm EW}$, the blue dashed line to
$T_{\rm EW} > T > T_{\rm QCD}$, and the green dot-dashed line to $T_{\rm QCD}>T$.
{\it Lower panel:} difference $\lambdabbar=\lambda-\Omega_{\rm vac0}$.
}
\label{fig_lambda-phi-rad}
\end{figure}

We found numerically that the system goes through the phase transitions more easily if the scalar
field difference $\lambdabbar$ is not too small as compared with the radiation density.
Indeed, from Eq.(\ref{eq:lambdabbar-pt}) $\lambdabbar$ jumps upward at the transition by
$\alpha_{\rm pt} \Omega_{\gamma}/(A a)^4$. If this is too large as compared with the previous
value of $\lambdabbar$, this may destabilize the system and lead to a vanishing Hubble expansion rate.
This means that the downward jump of the vacuum energy density in the Friedmann equation
(\ref{eq:EE00-hat}) is too strong and too fast to be quickly absorbed by the scalar field $\lambda$;
this yields a strong deviation of the Hubble expansion rate.
Although this problem may be cured by smoother phase transitions, or different kinetic functions
$K$, we can still find well-behaved solutions by ensuring that $\lambdabbar$ is not too small.
As we start the radiation-era solution (\ref{eq:scalar-rad}) at the end of the inflation era,
when $a \sim 10^{-28}$ and $T \sim 10^{15} \, {\rm GeV}$, much before the EW transition,
we require that $\lambdabbar$ does not decay much faster than the radiation component
$\tilde\rho_{\gamma}$.
From Eq.(\ref{eq:mu-lambda-rho-rad}) this corresponds to $2\sigma+\gamma-2 \simeq 0$.
On the other hand, it is convenient to have $\mu_\varphi \simeq 0$ as in Eq.(\ref{eq:mu-phi=0}),
so that the conformal factor $A$ does not evolve too much.
This is especially important at the time of the
Big Bang Nucleosynthesis (BBN), $T_{\rm BBN} \sim 1 \, {\rm MeV}$, as
the Hubble expansion rate of the Jordan frame must follow the standard radiation
era evolution to recover the usual abundance of primordial elements.
This implies
\beq
\left| \frac{d\ln A}{d\ln a} \right| \lesssim 10^{-2} \;\;\; \mbox{at} \;\;\; T_{\rm BBN} \sim 1 \, {\rm MeV} ,
\eeq
to make sure that standard predictions are not modified by more than a percent.
Both constraints are satisfied for
\beq
\gamma \simeq \frac{1}{3} \;\;\; \mbox{and} \;\;\;
\sigma \simeq \frac{5}{6} ,
\label{eq:gamma-nu-numerical}
\eeq
where the line $\mu_{\varphi}=0$ crosses the upper boundary $\sigma=1-\gamma/2$
of the allowed region (\ref{eq:gamma-rad-1}).
(In practice we take $\sigma$ slightly below $1-\gamma/2$ to be safely within
the allowed region).
For simplicity, we consider this solution for our numerical computation
(with also $\nu_A=1$, but this parameter has no physical effect and only corresponds to a choice
of normalization for $\varphi$).
However, more general functions such that $\gamma$ and $\sigma$ vary slowly
in the domains (\ref{eq:gamma-rad-1})-(\ref{eq:gamma-rad-2}),
with $\sigma \simeq 1 - \gamma/2$ until the EW transition and $\sigma \simeq 5/2-5\gamma$ around
the BBN, would also satisfy our requirements.

We show in Fig.~\ref{fig_lambda-phi-rad} the evolution with time of the scalar fields $\varphi$
and $\lambda$.
The three different line styles correspond to three successive epochs,
i) after the inflation era until the EW transition, ii) between the EW and QCD transitions,
and iii) after the QCD transition.
Between the phase transitions, $\varphi$ is roughly constant, because $\mu_\varphi \simeq 0$,
while the difference $\lambdabbar$ decays slightly faster than $\tilde\rho_\gamma$.
At each phase transition, $\varphi$ and $\lambdabbar$ jump along with the jump of the vacuum
energy density, and next recover the relaxing solution (\ref{eq:scalar-rad}).

\begin{figure}
\begin{center}
\epsfxsize=8. cm \epsfysize=6 cm {\epsfbox{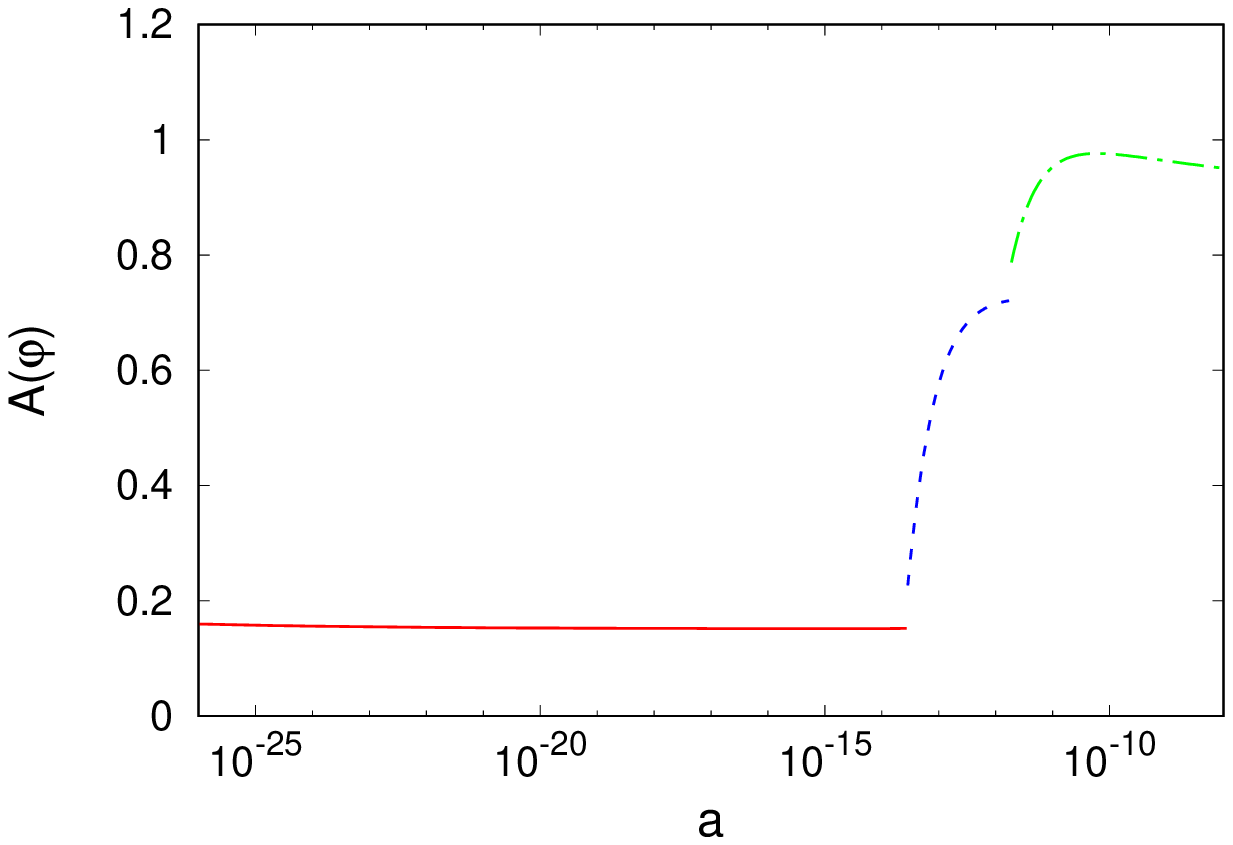}}
\epsfxsize=8. cm \epsfysize=6 cm {\epsfbox{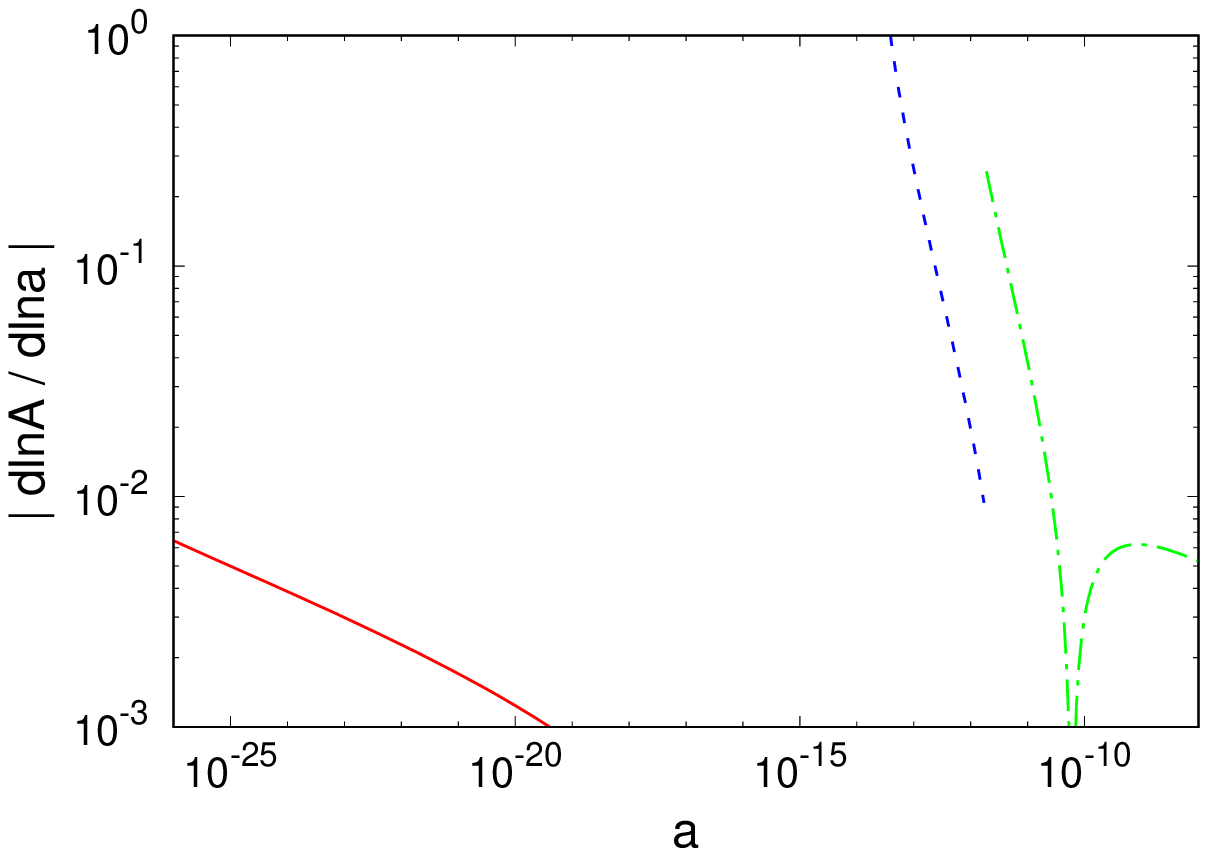}}
\end{center}
\caption{
{\it Upper panel:} conformal factor $A(\varphi)$ during the radiation era.
{\it Lower panel:} derivative $d\ln A/d\ln a$.
}
\label{fig_A-dlnA-rad}
\end{figure}

We show in Fig.~\ref{fig_A-dlnA-rad} the conformal factor $A$ and its logarithmic derivative with respect
to the scale factor.
The factor $A$ follows the evolution of $\varphi$, remaining almost constant between transitions
and jumping at the phase transitions.
This gives a time derivative $d\ln A/d\ln a$ that is of the order of $1\%$ in the smooth relaxation regime,
with jumps to high values at the transitions.
In particular, we can check that $|d\ln A/d\ln a| \leq 1\%$ at the BBN, which corresponds to
$a_{\rm BBN} \sim 10^{-11}$.
This ensures that the standard predictions for the primordial elements abundances are recovered to
$1\%$.

\begin{figure}
\begin{center}
\epsfxsize=8. cm \epsfysize=5.7 cm {\epsfbox{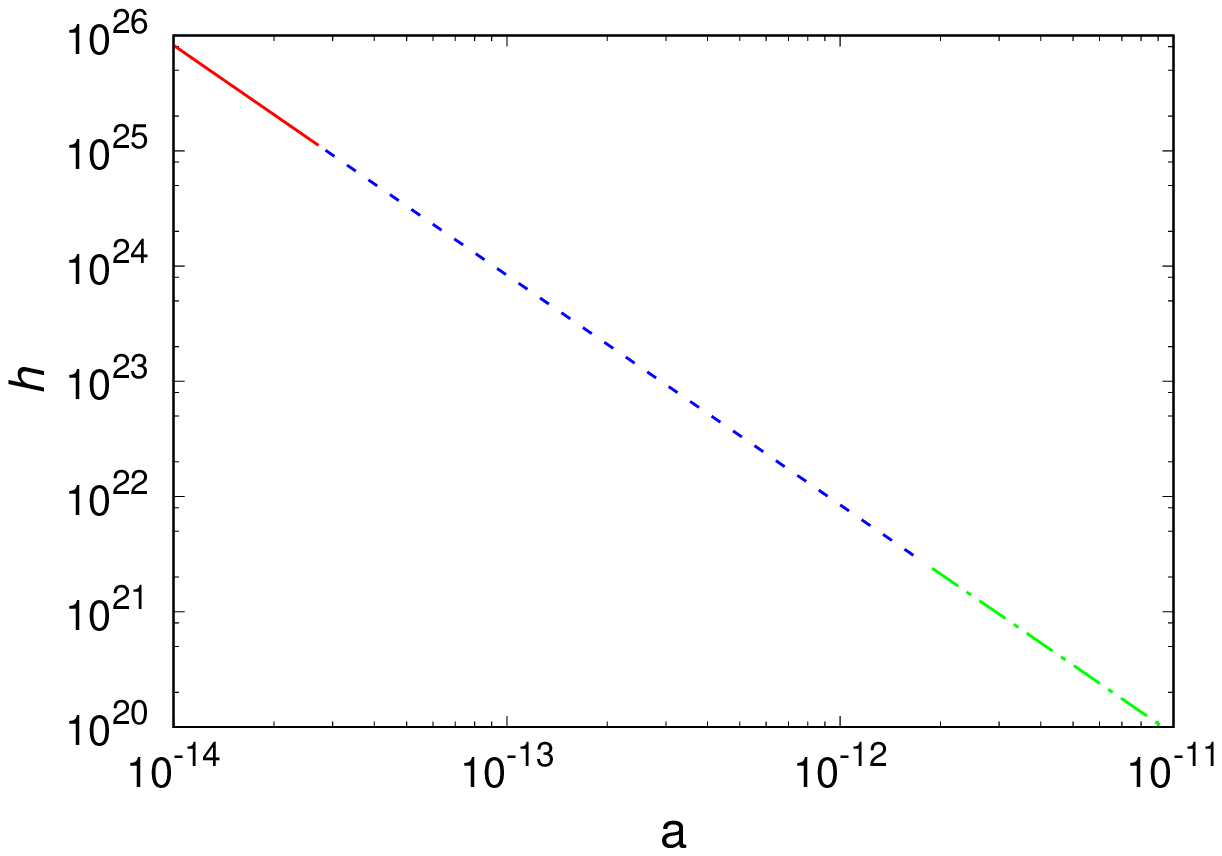}}
\epsfxsize=8. cm \epsfysize=5.7 cm {\epsfbox{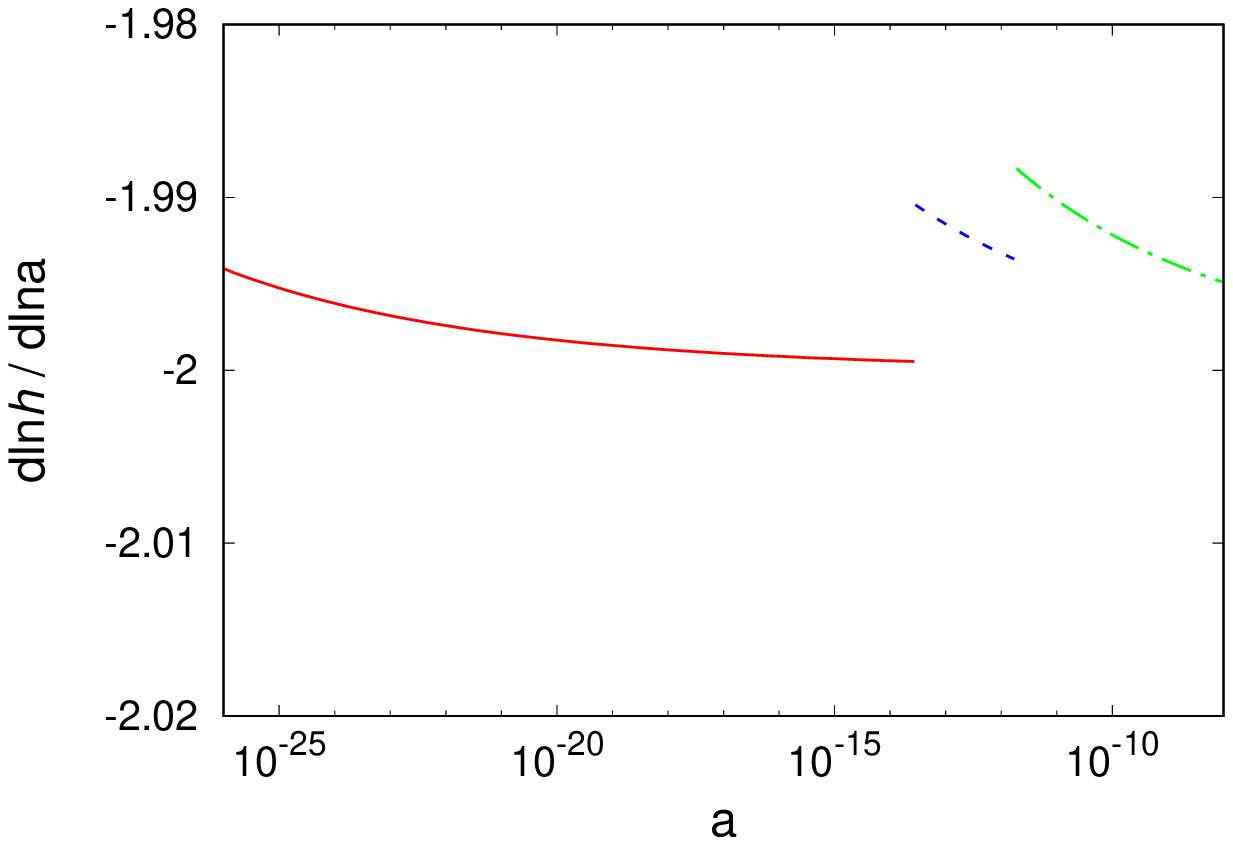}}
\epsfxsize=8. cm \epsfysize=5.7 cm {\epsfbox{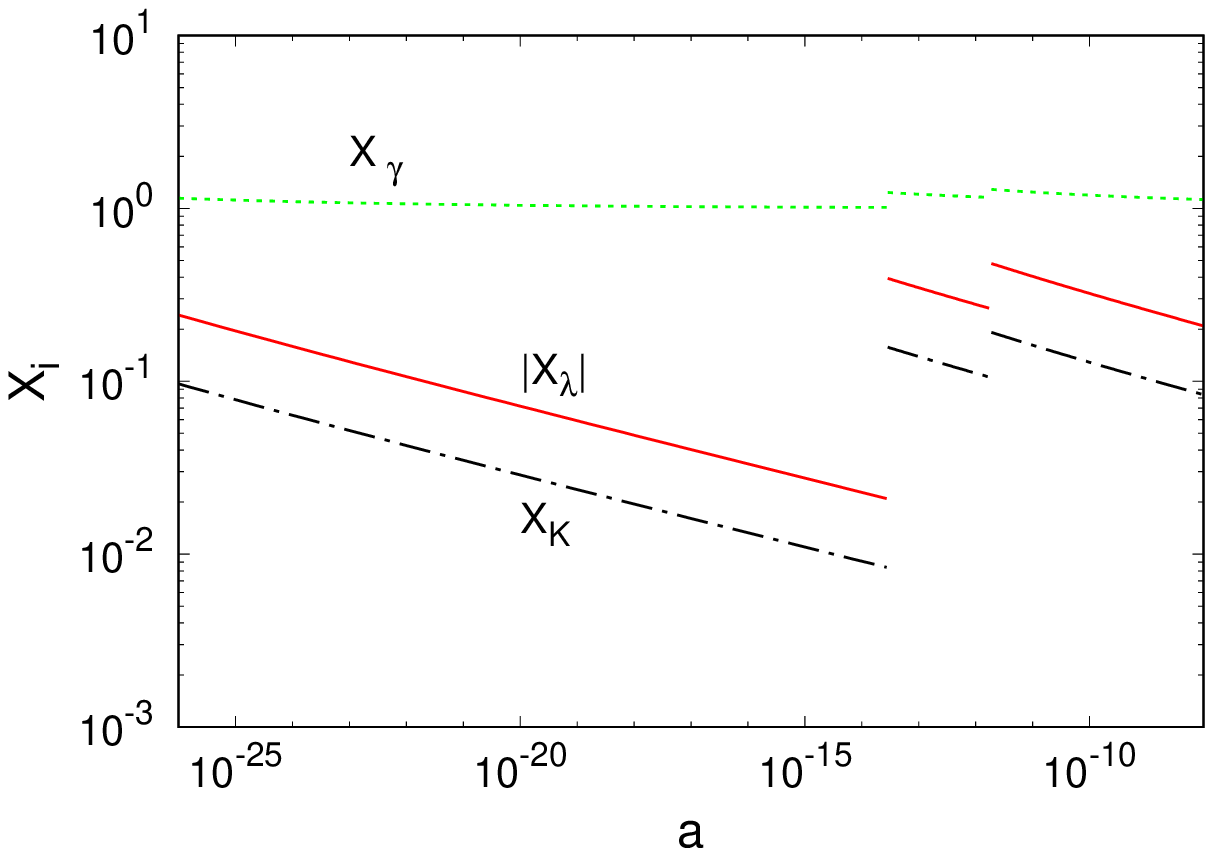}}
\end{center}
\caption{
{\it Upper panel:} reduced Hubble expansion rate around the phase transitions.
{\it Middle panel:} derivative $d\ln \hh/d\ln a$ during the radiation era.
{\it Lower panel:} contributions $X_i$ of Eq.(\ref{eq:X-def}) to the Friedmann equation.
}
\label{fig_dlnh-Xi-rad}
\end{figure}

We show in Fig.~\ref{fig_dlnh-Xi-rad} the reduced Hubble expansion rate 
(zooming around the phase transitions), its logarithmic derivative with respect to the scale 
factor, and the various contributions $X_i$ to the Friedmann equation
(\ref{eq:EE00-hat}), where we defined
\beqa
&& X_\lambda = \frac{A^4 (\Omega_{\rm vac0}-\lambda)}{\hh^2} , \;\;\;
X = \frac{A \Omega_0}{a^3 \hh^2} , \;\;\;
X_{\gamma} = \frac{\Omega_{\gamma 0}}{a^4 \hh^2} , \nonumber \\
&& X_K = \frac{(2\gamma-1) K}{\hh^2} .
\label{eq:X-def}
\eeqa
The Hubble expansion rate is continuous and deviations from the standard 
radiation-era decrease $\hh \propto a^{-2}$ cannot be distinguished in the upper panel.
Indeed, the middle panels shows that the time derivative $d\ln \hh/d\ln a$ remains 
very close to $-2$, with very small jumps at the phase transitions.
In particular, at the time of the BBN, we have $|2 + d\ln \hh / d\ln a| \simeq 1 \%$,
so that the standard BBN predictions are recovered within about $1\%$.

The contributions $X_\lambda$ and $X_K$ to the Friedmann equation, associated with the scalar
field $\lambda$, decay with time between phase transitions, as we verify the constraint
(\ref{eq:mu-lambda-rho-rad}). This decrease is very slow because we choose the coefficient
$\sigma$ as in Eq.(\ref{eq:gamma-nu-numerical}).
At each phase transition, where $\tilde{V}_{\rm vac}$ jumps by an amount of the order of the
radiation density $\tilde\rho_{\gamma}$, the contribution $X_\lambda$ jumps to a value of order unity.
In agreement with Eq.(\ref{eq:K-A-lambda}), the kinetic energy of the scalar field shows a similar jump.
The sum of these contributions is negative; this implies a small positive jump for the contribution
$X_{\gamma}$ of the radiation component.

Because the EW transition occurs much later than the beginning of the radiation era,
the difference $\lambdabbar$ has had time to decay much below the radiation density.
This leads to a strong jump for $\lambdabbar$ at the EW transition.
This also yields the large jumps seen in Figs.~\ref{fig_lambda-phi-rad} and
\ref{fig_A-dlnA-rad} for $\varphi$ and $A$.
In contrast, the QCD transition occurs shortly after the EW transition and the difference
$\lambdabbar$ has not yet decayed much below $\tilde\rho_\gamma$.
This leads to a smaller jump for $\lambdabbar$ and $\varphi$ at this second transition.

We can also check that the nonrelativistic matter density remains negligible at all times shown
in Fig.~\ref{fig_dlnh-Xi-rad}.

\begin{figure}
\begin{center}
\epsfxsize=8. cm \epsfysize=6 cm {\epsfbox{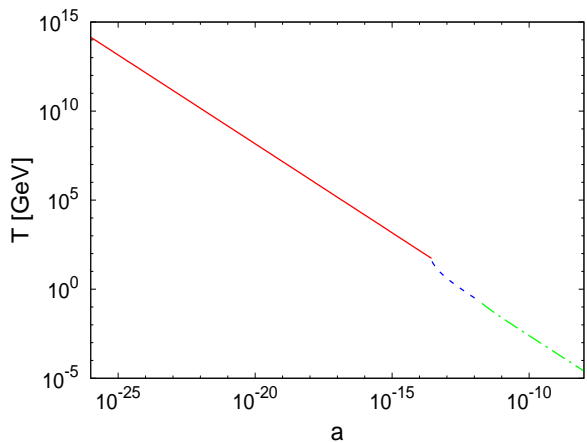}}
\end{center}
\caption{
Temperature of the radiation component during the radiation era.
}
\label{fig_T-rad}
\end{figure}

We show in Fig.~\ref{fig_T-rad} the temperature of the radiation component in the Jordan frame,
which we define by $T^4=\tilde\rho_\gamma$.
Between transitions it decays as $1/\tilde{a}$, which closely follows $1/a$ as $A$ is almost constant.
It decreases somewhat faster right after the EW transition because of the sudden increase of $A$.

\section{Matter era}
\label{sec:Matter-era}

\subsection{Impact of the coupling to matter}
\label{sec:impact-coupling-matter}

We have seen in the previous section that the radiation era is easily recovered, with a cancellation
of the vacuum energy jumps at the EW and QCD phase transitions.
This is because the conformal coupling $A(\varphi)$ only couples the scalar fields to the vacuum
and matter energy densities, through the right-hand side in Eq.(\ref{eq:dS-dphi-hat}).
The coupling to the vacuum energy density gives rise to the cancellation mechanism we wish to achieve,
while the coupling to nonrelativistic matter is irrelevant during the radiation era as it is a negligible
component.
However, at later times we must recover the matter era, where nonrelativistic matter is the dominant
component of the Universe.
This is more difficult as the coupling generated by the right-hand side in Eq.(\ref{eq:dS-dphi-hat})
would typically mean that one fourth of the matter density is now canceled by the scalar field.

This difficulty to recover the matter era is actually common with some other self-tuning models;
see for instance the discussion in \cite{Appleby:2018yci}.
It arises from the fact that at a given time there is no simple and unambiguous way to distinguish
between the vacuum and the matter energy densities.
In the sequestering model \cite{Padilla_2014,Kaloper:2014dqa}, this problem is solved
in a simple and elegant fashion by
the use of global variables. Then, the value of the cancellation field $\lambda$ is set by an
integral over all spacetime of the trace of the energy-momentum tensor, and the integral is naturally
dominated by the contribution of the vacuum energy density at late times (while the matter component
is diluted by the expansion of the Universe).
In the dynamical model that we develop in this paper, we cannot use this remedy and we must face
the consequences of the coupling to the trace of the energy-momentum tensor at each cosmological
time.

One could try to cancel the matter component in the right-hand side in Eq.(\ref{eq:dS-dphi-hat})
by a component of the kinetic term on the left-hand side. For instance, we considered
kinetic functions of the form $K=K_X e^{\nu\varphi} X^{\gamma} + K_Z e^{\nu_Z\varphi} Z$,
where we add a $Z$-component to the form (\ref{eq:K-rad}).
Then, we obtained solutions such that the $K_Z$ terms, associated with $\varphi$,
cancel the $\Omega_0$ term in Eq.(\ref{eq:dS-dphi-hat}), and $\lambdabbar$ shows a fast decay.
Unfortunately, these solutions are not stable and the system typically converges to another
solution where $\lambdabbar$ is constant while $\varphi$ runs towards $-\infty$
so that the coupling $A$ decreases with time.
This can be easily understood from the form of the equations of motion
(\ref{eq:dS-dphi-hat})-(\ref{eq:dS-dlambda-hat}).
If the first equation (\ref{eq:dS-dphi-hat}) mainly governs $\varphi$, through the $K_Z$ terms,
so that it cancels the matter terms, it also means that it does not dictate $\lambda$
(a single equation does not simultaneously govern two fields).
Then, $\lambda$ is set by the second equation (\ref{eq:dS-dlambda-hat}), which only depends on
derivatives of $\lambda$. Therefore, it always admits a constant solution, which is typically
more stable. This quickly makes the $\lambda$ component greater than the matter component
in the Friedmann equations, and we escape from the matter era.
We did not conduct a thorough investigation of this scenario, and it may happen that more
complex kinetic functions provide a stable and fast decay of $\lambdabbar$.
On the other hand, it may be a clue that the matter era is only a transient between the radiation
and dark energy eras, which could point towards such scenarios.

In this paper, we consider instead solutions where $\lambdabbar$ scales as a constant fraction
of the matter component. This provides an Hubble expansion rate that obeys the usual matter era
scaling $\hh \propto a^{-3/2}$, but with a proportionality factor that is typically different than in
the $\Lambda$CDM cosmology, because of the contribution from the scalar fields.

\subsection{Solutions driven by the matter}
\label{sec:scaling-matter}

As we keep the same form (\ref{eq:K-rad}) for the kinetic function as in the radiation era
(but we allow the parameters to be different), the equations of motion
(\ref{eq:dS-dphi-rad})-(\ref{eq:dS-dlambda-rad}) still apply.
Again, we first present an analytic study of the relevant solutions and their
linear stability, to obtain the range of the parameters $\nu_X$ and $\gamma$ that lead to
the desired properties.
Since we aim at recovering the matter era expansion, to be consistent with observational data,
we now write the Hubble expansion rate as
\beq
\hh = h_\star e^{-3\eta/2} .
\label{eq:h-2eta-matter}
\eeq
Moreover, we require the Planck mass to be constant in the Jordan frame. Then, the
conformal factor $A(\varphi)$ must remain almost constant with time.
This means that the scalar field $\varphi$ must also remain almost constant.
As we explained above, the scalar field $\lambdabbar$ must also follow the matter component
$\Omega_0/a^3$ in Eq.(\ref{eq:dS-dphi-rad}). Indeed, if $\lambdabbar$ becomes much greater
it will dominate in the Friedmann equation (\ref{eq:EE00-hat}) and we do not recover the matter era.
On the other hand, if $\lambdabbar$ becomes much smaller it can be neglected in
Eq.(\ref{eq:dS-dphi-rad}). Then, the equations of motion (\ref{eq:dS-dphi-rad})-(\ref{eq:dS-dlambda-rad})
only depend on derivatives of $\lambdabbar$ and we typically branch to a constant-$\lambdabbar$
solution, which will eventually take over the matter component.
Therefore, we now look for solutions of the form
\beq
\varphi = \varphi_\star , \;\;\;
\lambdabbar =  \lambda_\star e^{-3\eta} .
\label{eq:scalar-matter}
\eeq
Substituting into the equations of motion (\ref{eq:dS-dphi-rad})-(\ref{eq:dS-dlambda-rad}), we obtain
the three constraints
\beqa
&& \gamma= \frac{1}{3} , \;\;\;\;
\sigma = 1 - \frac{\Omega_0}{4\lambda_\star A_\star^3} e^{-3\nu_A\varphi_\star} ,
\nonumber \\
&& K_X  e^{\nu_X\varphi_\star}   \left( 9 h_\star^2 \lambda_\star^2/2 \right)^{1/3}
= - \lambda_\star A_\star^4 e^{4\nu_A\varphi_\star} .
\label{eq:matter-contraint-3}
\eeqa
As compared with the analysis of the radiation era in (\ref{eq:rad-contraint-3}),
we have now imposed $\mu_\lambda=-3$ and $\mu_\varphi=0$.
This uniquely determines the exponent $\gamma=1/3$ while the last two equations
in (\ref{eq:matter-contraint-3}) set the normalization $\varphi_\star,\lambda_\star$ of the
solution (\ref{eq:scalar-matter}), which is no longer defined up to a fixed rescaling.
This is because the $\Omega_0$ term in Eq.(\ref{eq:dS-dphi-rad}) provides an external source
that governs the amplitude of the scalar fields.
We must again require that this solution be stable.
Therefore, we now study the evolution of the perturbations $\delta\varphi, \delta\lambdabbar$ and
$\delta\hh$ at linear order, with
\beqa
&& \varphi = \varphi_\star + \delta\varphi , \;\;\;
\lambdabbar =  \lambda_\star e^{-3\eta} (1+\delta\lambdabbar) , \nonumber \\
&& \hh = h_\star e^{-3\eta/2} (1+\delta \hh) .
\label{eq:scalar-matter-delta}
\eeqa
We must now take into account the perturbation of the Hubble expansion rate,
as the scalar fields give a non-negligible contribution to the Friedmann equation that scales
like the matter density.
The Friedmann equation (\ref{eq:EE00-hat}) and the constraint equation (\ref{eq:dS-dphi-rad})
give $\delta \hh$ and $\delta\varphi$ in terms of $\delta\lambdabbar$.
Substituting into Eq.(\ref{eq:dS-dlambda-rad}) we obtain
\beq
\frac{d^2\delta\lambdabbar}{d\eta^2} + \frac{3}{2} \frac{d\delta\lambdabbar}{d\eta}
+ \frac{54 (2-5\sigma+3\sigma^2)}{3+\sigma-12\sigma^2} \delta\lambdabbar = 0 .
\label{eq:decay-matter}
\eeq
This gives two decaying modes when
\beq
\frac{1-\sqrt{145}}{24} < \sigma < \frac{1+\sqrt{145}}{24} \;\;\; \mbox{or} \;\;\; \frac{2}{3} < \sigma < 1 .
\label{sec:sigma-bound-matter}
\eeq
On the other hand, the relative contribution of the scalar fields to the Friedmann equation
(\ref{eq:EE00-hat}) reads as
\beq
X_\lambda + X_K = \frac{1}{6 \sigma - 5} ,
\label{eq:Friedmann-matter-era-scalar}
\eeq
while the requirement $\hh^2>0$ implies
\beq
\hh^2 > 0 : \;\;\; \sigma < \frac{5}{6} \;\;\; \mbox{or} \;\;\; \sigma > 1 .
\label{eq:h2-positive}
\eeq

To be consistent with observations, the contribution of the scalar fields to the Friedmann equation
should be small, which points to small values of $\sigma$.
In fact, even for $\sigma \simeq -0.46$, which corresponds to the lower boundary in
(\ref{sec:sigma-bound-matter}), the scalar fields contribute for $13\%$ to $\hh^2$,
which is most likely too large to obey observational constraints.
Another shortcoming is that we need to change the form of the kinetic function between the
radiation and matter eras.
Indeed, while $\gamma = 1/3$ can be kept identical for both the radiation era,
from the constraint (\ref{eq:gamma-rad-1}), and the matter era, from the first constraint
in (\ref{eq:matter-contraint-3}), the exponent $\sigma$ must decrease from about $5/6$ to
about $-0.4$ (if we wish to minimize the contribution to the Friedmann equation).
This change can start somewhat before the matter era, but should not occur too early after the
last matter phase transition as a small value of $\sigma$
would trigger instabilities, being outside of the stability range (\ref{eq:gamma-rad-1}).
This corresponds to some degree of tuning, in the sense that this change of the kinetic function
appears as a coincidence, unless the scalar field Lagrangian ``knows'' about the matter Lagrangian
and the final background radiation and matter densities.
This means that our simple example, based on the kinetic function (\ref{eq:K-rad}),
is not very satisfactory.
Some other self-tuning models introduced to tackle the cosmological constant problem
also share this behavior. For instance, in the self-tuning models presented in
\cite{Charmousis:2011bf,Copeland:2012qf}, the radiation and matter eras also correspond
to different terms in the Lagrangian being dominant.
It would be desirable to find a kinetic function that can simultaneously reproduce the
radiation and matter eras and also give a small enough contribution from the scalar fields
to the Friedmann equation.
We leave this investigation for future works.

\subsection{Numerical computation}
\label{sec:matter-numerical-1}

\begin{figure}
\begin{center}
\epsfxsize=8. cm \epsfysize=6 cm {\epsfbox{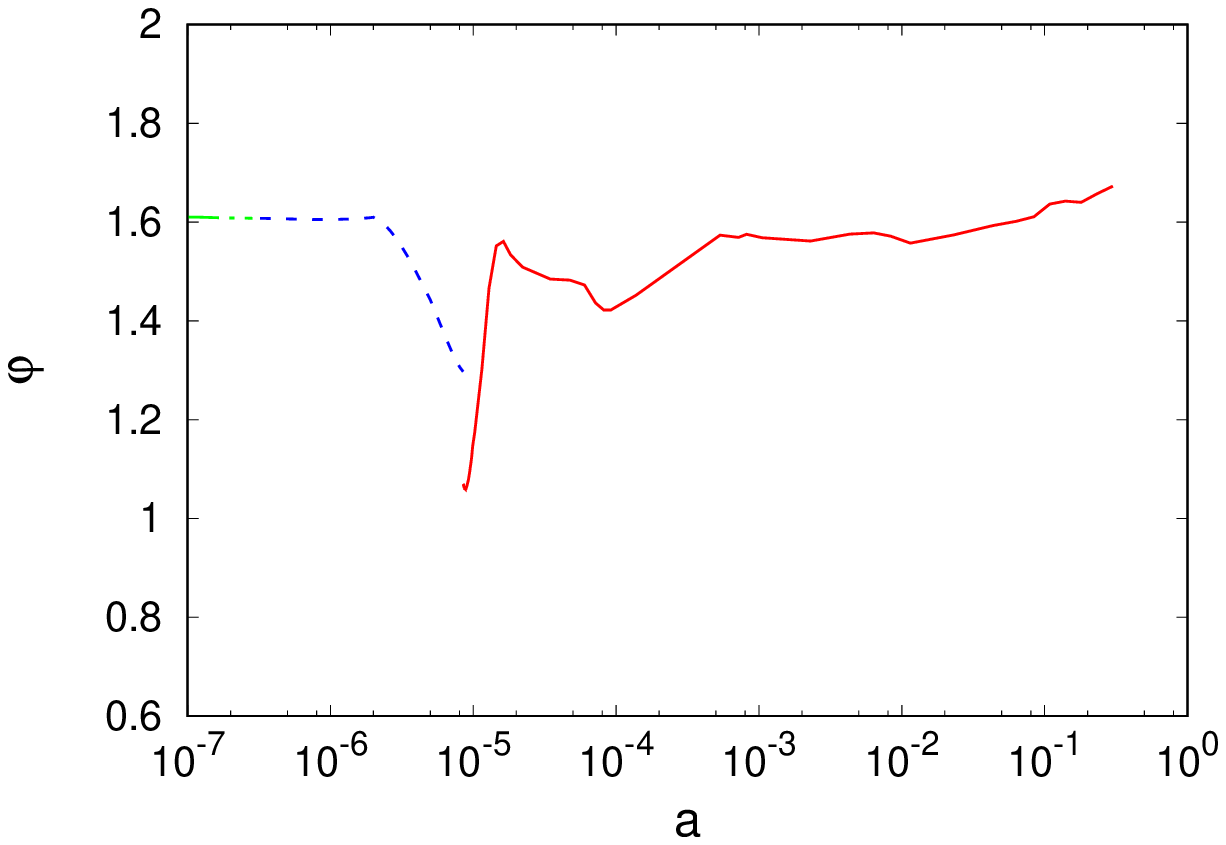}}
\epsfxsize=8. cm \epsfysize=6 cm {\epsfbox{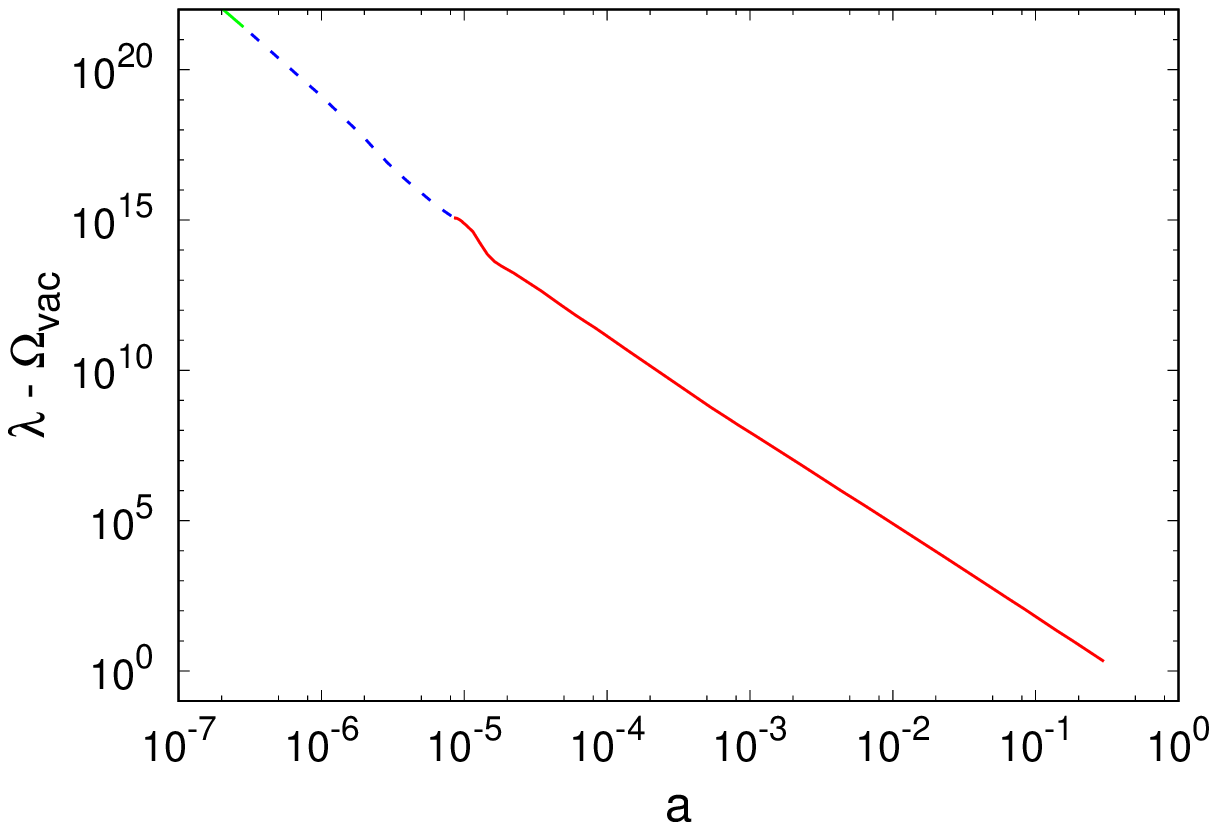}}
\end{center}
\caption{
{\it Upper panel:} scalar field $\varphi$ during the late radiation era and the matter era.
{\it Lower panel:} difference $\lambdabbar=\lambda-\Omega_{\rm vac0}$.
}
\label{fig_lambda-phi-matter}
\end{figure}

\begin{figure}
\begin{center}
\epsfxsize=8. cm \epsfysize=6 cm {\epsfbox{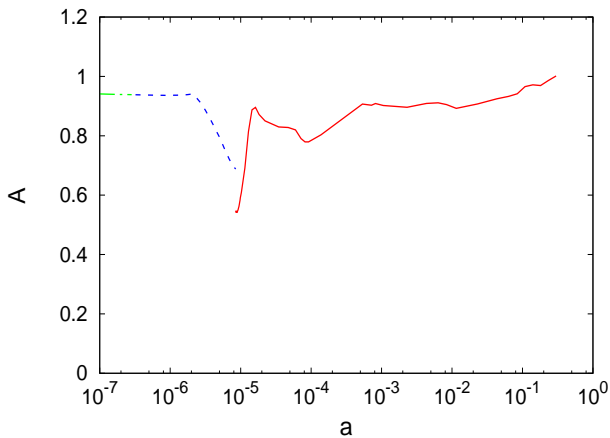}}
\end{center}
\caption{
Conformal factor $A(\varphi)$ during the late radiation era and the matter era.
}
\label{fig_A-matter}
\end{figure}

We now present an explicit numerical implementation of the solutions found in the previous section.
Our implementation of the decrease of the exponent $\sigma=\nu_X/(\nu_A)$ of the kinetic function,
from its radiation-era value $5/6$ down to its final matter era value $-0.4$, which we choose close to
the lower boundary (\ref{sec:sigma-bound-matter}), is illustrated
by the upper panel in Fig.~\ref{fig_lambda-phi-matter}, which shows the evolution with time
of the scalar field $\varphi$.
The three line styles correspond to three different stages.
The green dot-dashed line is the end of the radiation era, already displayed in Fig.~\ref{fig_lambda-phi-rad},
with the kinetic function of section~\ref{sec:radiation-numerical-1}.
The blue dotted line corresponds to a slow and smooth decrease of the exponent $\sigma$.
During the first flat part, we decrease $\sigma$ from $5/6$ down to $0.75$ along
the line $\mu_\varphi=0$ of Eq.(\ref{eq:mu-phi=0}), so that $\varphi$ remains constant.
Next, we further decrease $\sigma$ down to $0.6$ while keeping $\gamma$ below $0.37$.
This falls below the line defined by Eq.(\ref{eq:mu-phi=0}), so that $\varphi$ is no longer constant
and decreases.
Next, the solid line starting at $a \sim 10^{-5}$ starts with a discontinuous jump of $\sigma$ down to $0.5$,
to reach the lowest allowed range (\ref{sec:sigma-bound-matter}) within the basin of attraction of the solution
(\ref{eq:scalar-matter}). We found numerically that using instead a slow and continuous transition
down to $\sigma=0.5$ makes it difficult to reach the solution (\ref{eq:scalar-matter}) and leads
to strong instabilities, in agreement with the forbidden range $0.54 \lesssim \sigma \lesssim 0.66$
found in (\ref{sec:sigma-bound-matter}).
(Because these events take place somewhat before the radiation-matter equality,
these bounds do not rigorously apply but are suggestive of possible problems.)
Next, we slowly decrease $\sigma$ down to $-0.4$ in a continuous manner, while $\gamma$ goes
to $1/3$, so as to minimize the contribution (\ref{eq:Friedmann-matter-era-scalar}) of the scalar fields
to the Friedmann equation.
We tune the speed of this last step so that the final value of $\varphi$ is almost equal to the
one obtained during the radiation era after the QCD transition. This ensures that after these steps
the Planck mass remains equal to its value at the BBN.
The transitions associated with the decrease of $\sigma$ during the matter era lead to small oscillations.
This agrees with the fact that the roots of Eq.(\ref{eq:decay-matter}) have a nonzero imaginary part, and
a real part equal to $-3/4$. This corresponds to oscillatory decaying modes, with an envelope
that only falls as $a^{-3/4}$.
Some of these oscillations may disappear by using a continuous kinetic function,
whereas in our numerical implementation we discretize the change of $\sigma$ as a series
of small jumps, while ensuring that the junction conditions are satisfied across each transition.
We provide more details of our numerical procedure in appendix~\ref{app:jumps}.

Thus, we consider a scenario where the kinetic function $K(\varphi;X,Y,Z)$ takes the simple form
(\ref{eq:K-rad}) in both the radiation and matter eras, but where the parameters
$K_X, \nu_X$ and $\gamma$ are different and evolve with the cosmic time.
This change of the kinetic function is possible, and does not imply a multivalued function,
because $X$ shows a monotonic decrease with time, through the radiation and matter era,
along with the Hubble expansion rate $\hh$ and the scalar field derivative $d\lambdabbar/d\eta$.
Therefore, we can use $X$ as a "clock" and consider that the different forms of the kinetic
functions correspond to different ranges of its argument $X$.
A more realistic scenario would use a more complex kinetic function, which smoothly interpolates
between these different regimes.

\begin{figure}
\begin{center}
\epsfxsize=8. cm \epsfysize=6 cm {\epsfbox{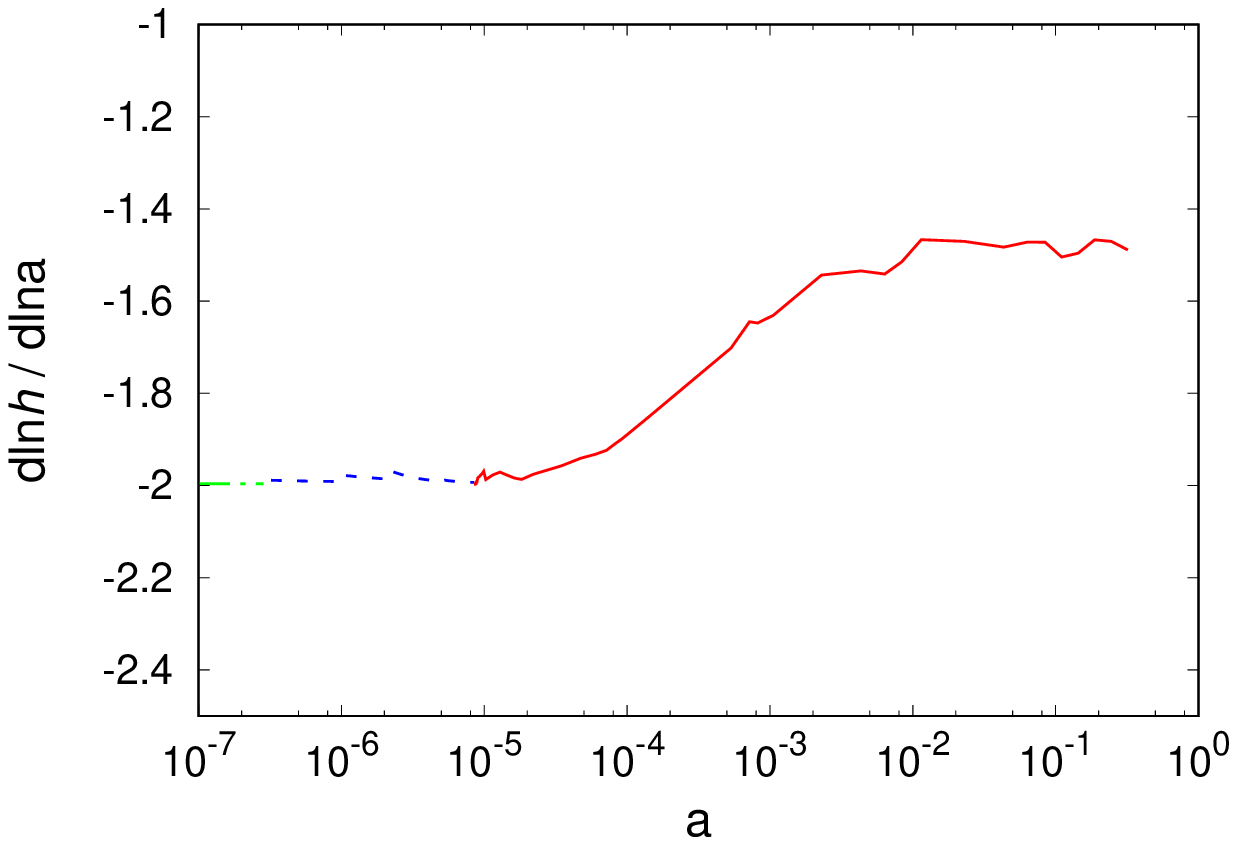}}
\epsfxsize=8. cm \epsfysize=6 cm {\epsfbox{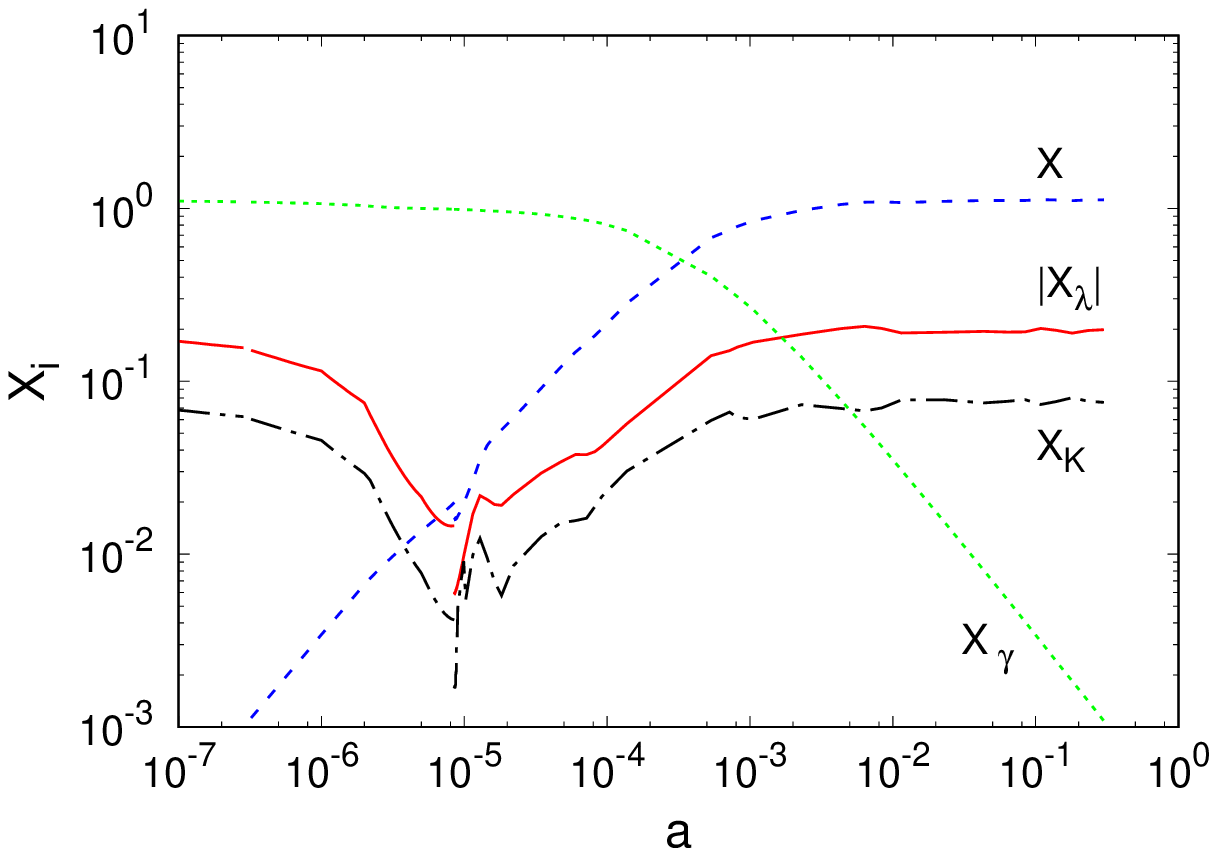}}
\end{center}
\caption{
{\it Upper panel:} derivative $d\ln \hh/d\ln a$ during the late radiation era and the matter era.
{\it Lower panel:} contributions $X_i$ to the Friedmann equation.
}
\label{fig_dlnh-Xi-matter}
\end{figure}

As seen in the lower panel in Fig.~\ref{fig_lambda-phi-matter}, $\lambdabbar$ keeps decreasing
with time. While during the radiation era it decreased roughly as $a^{-4}$,
because of our choice (\ref{eq:gamma-nu-numerical}), during the matter era it tracks the
matter density and decreases as $a^{-3}$, as in Eq.(\ref{eq:scalar-matter}).

The coupling function $A(\varphi)$ remains given by Eq.(\ref{eq:A-rad-def}) throughout.
As seen in Fig.~\ref{fig_A-matter}, this conformal factor follows the evolution of $\varphi$ and
is roughly constant during the matter era. In practice, we normalize $\varphi$ at the beginning
of the radiation era, hence $A$, so that the asymptotic value reached at the end of the matter era
is unity. This means that at low redshifts the Einstein-frame and Jordan-frame scale factors
and Planck masses are equal, as well as the Hubble expansion rate. Therefore, $M_{\rm Pl}$
and $H_0$ are simply given by their observed values.
In particular, because the transition of the kinetic function occurs somewhat before the radiation-matter
equality, the fields have relaxed before the time of the last-scattering surface ($a \sim 10^{-3}$)
probed by the Cosmic Microwave Background (CMB) anisotropies.
This ensures that $M_{\rm Pl}$ has remained almost constant since the time of the last scattering
and that we recover the standard statistics of the CMB, provided the background expansion
follows the standard $\Lambda$CDM expansion at later times, or that we recover the same angular
distances.
We enforce a small increase of $\varphi$ and $A$ at the end of the matter era, by decreasing slightly
the exponent $\gamma$, to authorize the transition to the dark energy era as described in
section~\ref{sec:Dark-energy-era} below.

We show in Fig.~\ref{fig_dlnh-Xi-matter} the logarithmic derivative of the Hubble expansion rate
with respect to the scale factor and the various contributions $X_i$ to the Friedmann equation
(\ref{eq:EE00-hat}), which were defined in (\ref{eq:X-def}).
The time derivative $d\ln h/d\ln a$ goes from $-2$, which corresponds to the radiation
era, to $-1.5$, which corresponds to the matter era.
The small oscillations are due to the oscillations of the scalar field $\lambda$, which yields a contribution
to the Friedmann equation that is not completely negligible.
In agreement with the analysis of section~\ref{sec:scaling-matter} and
Eq.(\ref{eq:Friedmann-matter-era-scalar}), the contributions $X_\lambda$
and $X_K$ to the Friedmann equation, associated with the scalar
field $\lambda$, converge to a constant fraction of the matter contribution in the matter era.
For our choice $\sigma=-0.4$ this gives $X_\lambda+X_K \simeq -0.14$.

Beyond the background cosmology level, cosmological perturbations will also be  affected 
by the presence of the fields $\lambda$ and $\varphi$. This could have an effect on CMB 
physics and the large-scale structures of the Universe. A detailed study of these issues 
is left for future works.

\section{Dark energy era}
\label{sec:Dark-energy-era}

\subsection{End of the cancellation mechanism}
\label{sec:end-cancellation}

Eventually, we must exit from the matter era and recover the dark energy era
at current times. Again, this will correspond to a change of the kinetic
(and coupling) functions. However, contrary to the case of the exit from the
radiation era, this does not really involve an additional tuning, as compared
with the $\Lambda$CDM cosmology. Indeed, for the exit from the radiation era,
we had to introduce a shift of the kinetic function somewhat before the
radiation-matter equality. This can be seen as a coincidence between two
unrelated events (unless the scalar field Lagrangian ``knows'' about details of the
matter Lagrangian that governs the baryogenesis and the relic matter density).
In contrast, for the exit from the matter era, there is no coincidence with an
external event because the dark energy era will be generated by the change itself
of the scalar field functions; it is not an external
event associated with another component such as an external quintessence fluid.
However, we still face the standard coincidence problem associated with the question
of why this transition happens now, and not earlier or much further in the future.

Within our framework, which builds a cancellation mechanism of the vacuum energy
density through the scalar field $\lambda$, it is clear that dark energy eras,
or more precisely, epochs where the expansion is driven by an effective cosmological
constant, appear naturally as periods where this cancellation mechanism stops
or is ineffective. Because this mechanism is linked to the conformal coupling
$A(\varphi)$, acting as a Lagrange multiplier as explained above Eq.(\ref{eq:S-sym})
and in section~\ref{sec:cancellation-radiation}, this mechanism automatically stops
or becomes inefficient when $A(\varphi)$ becomes a constant, or $dA/d\varphi$
is negligible. Another possibility is to make the kinetic function large,
so that the right-hand side in the equation of motion
(\ref{eq:dS-dphi}) is negligible.
This again makes the conformal coupling inefficient.

In this paper, we consider the simple scenario where $dA/d\varphi$ becomes zero
at late times. Then, the equations of motion (\ref{eq:dS-dphi-hat})-(\ref{eq:dS-dlambda-hat})
only depend on derivatives of $\lambda$. This means that generically there exists
a solution with a constant $\lambda$, with a value that is set by the initial
conditions (i.e., just before the vanishing of $dA/d\varphi$).
Provided this solution is stable and $\lambdabbar \equiv \lambda - \Omega_{\rm vac0}$
is negative, it will play the role of a cosmological constant in the Friedmann equation
(\ref{eq:EE00-hat}).
On the other hand, the equation of motion (\ref{eq:dS-dlambda-hat}) shows that
if we wish to have $A$ and $\lambda$ being constant, we need $\partial K/\partial Y$
and $d\varphi/d\eta$ to be nonzero. This is related to the need to avoid Weinberg's theorem
as explained in section~\ref{sec:cancellation-radiation}: we need a time dependent
background. To have $\partial K/\partial Y \neq 0$, we simply consider the case
where at late times the kinetic function becomes
\beq
K(\varphi;X,Y,Z) = K_X  X^{\gamma} + K_Y Y , \;\;\; \gamma > 0 ,
\label{eq:K-DE}
\eeq
while the coupling function is constant and equal to unity
\beq
\varphi > \varphi_{\rm DE} : \;\;\; A(\varphi) = A_\star = 1 .
\label{eq:A-DE}
\eeq
Here we take a sharp transition, at a time $\eta_{\rm DE}$.
The coupling $A(\varphi)$ is continuous, as $A(\varphi)$ computed in the matter era
and displayed in Fig.~\ref{fig_A-matter} reaches unity at time $\eta_{\rm DE}$.
We also take a nonzero kinetic term $K_Y Y$ to appear shortly before $\eta_{\rm DE}$
while remaining subdominant, so as to play no role in the dynamics before
$\eta_{\rm DE}$. Thus, at time $\eta_{\rm DE}$ the exponent $\nu_A$ goes to zero
while the kinetic function goes from $K_{X_{\rm mat}} e^{\nu_{X_{\rm mat}} \varphi}
X^{\gamma_{\rm mat}} + K_{Y_{\rm mat}} Y$,
with $\nu_{X_{\rm mat}}=-1.6$ and $\gamma_{\rm mat}=1/3$ as in the end of
the matter era described in section~\ref {sec:matter-numerical-1},
to $K_{X_{\rm DE}} X^{\gamma_{\rm DE}} + K_{Y_{\rm DE}} Y$.
We take $\gamma_{\rm DE}=1/4$ and we ensure that the equations of motion
(\ref{eq:dS-dphi-hat})-(\ref{eq:dS-dlambda-hat}) are satisfied across the transition
by requiring continuity of $\frac{\partial K}{\partial Y} \frac{d\lambdabbar}{d\eta}$
and $\frac{\partial K}{\partial X} \frac{d\lambdabbar}{d\eta}
+ \frac{\partial K}{\partial Y} \frac{d\varphi}{d\eta}$.
We also require continuity of $\hh$ and $d\ln\hh/d\eta$.
These conditions set $K_{X_{\rm DE}}$ and $K_{Y_{\rm DE}}$ and also provide
$\frac{d\lambdabbar}{d\eta}$ and $\frac{d\varphi}{d\eta}$ just after the transition.

This transition is possible because the scalar field $\varphi$ can act as a ``clock''.
Indeed, we can see from Figs.~\ref {fig_lambda-phi-rad} and \ref{fig_lambda-phi-matter}
that $\varphi$ is greater at time $\eta_{\rm DE}$ than at all previous times.
This ensures that the transition to (\ref{eq:K-DE})-(\ref{eq:A-DE})
does not imply multivalued functions and is set by the crossing of the boundary value
$\varphi_{\rm DE}$.
In more realistic scenarios, the kinetic and coupling functions would show a smooth
transition, which would automatically ensure that $\hh$ and $d\ln \hh/d\eta$
are continuous. However, here we do not perform a complete study with an accurate
quantitative match with observational data, which we leave to future works.
We simply describe how a dark energy era can naturally occur at late times
within our framework.

\subsection{Numerical computation}
\label{sec:DE-numerical-1}

\begin{figure}
\begin{center}
\epsfxsize=8. cm \epsfysize=6 cm {\epsfbox{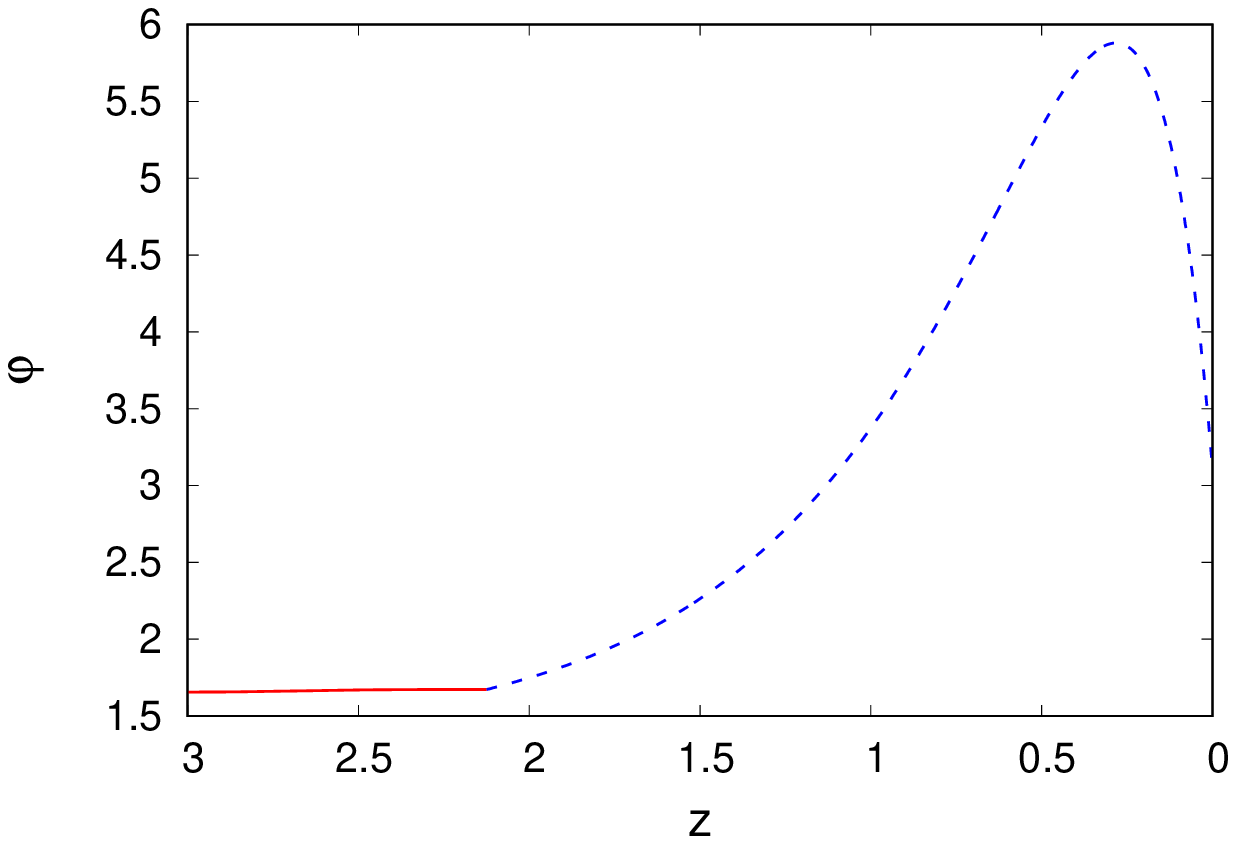}}
\epsfxsize=8. cm \epsfysize=6 cm {\epsfbox{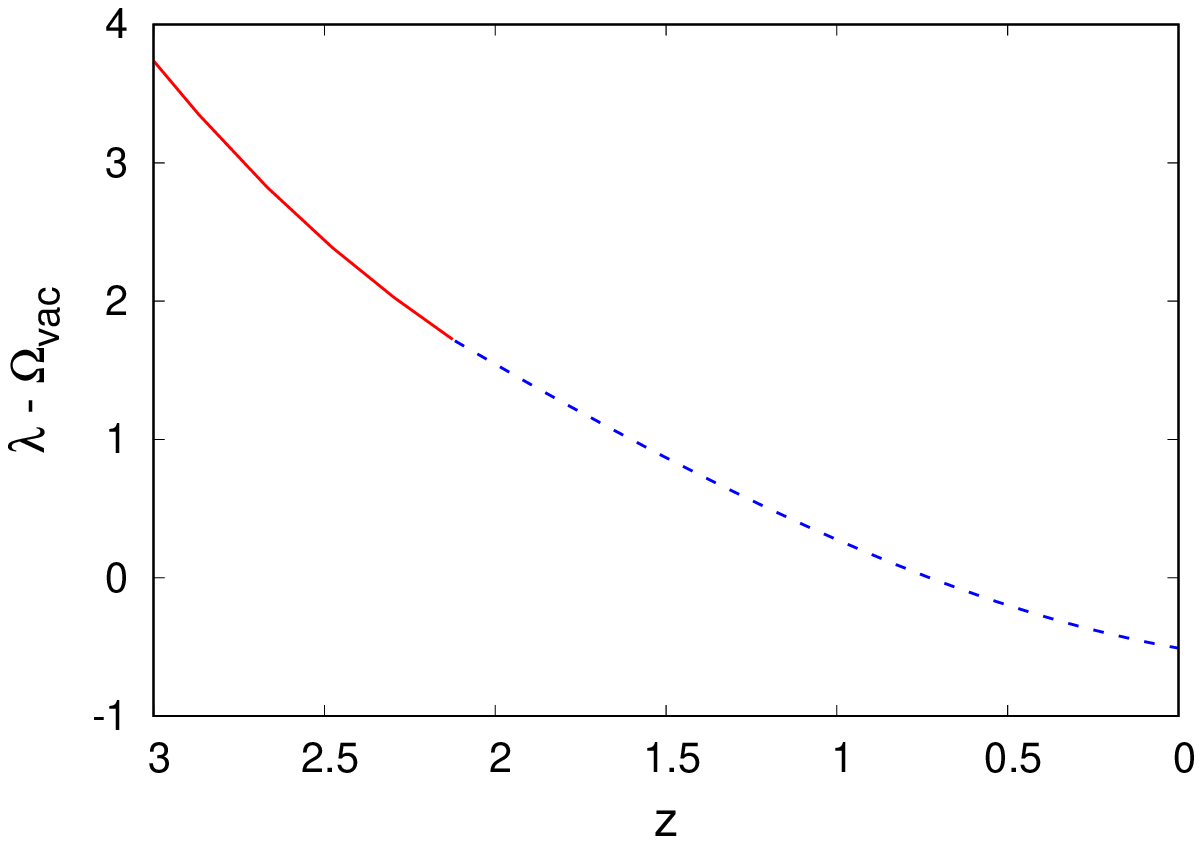}}
\end{center}
\caption{
{\it Upper panel:} scalar field $\varphi$ during the late matter era (red solid line)
and the dark energy era (blue dashed line), as a function of redshift.
{\it Lower panel:} difference $\lambdabbar=\lambda-\Omega_{\rm vac0}$.
}
\label{fig_lambda-phi-DE}
\end{figure}

\begin{figure}
\begin{center}
\epsfxsize=8. cm \epsfysize=6 cm {\epsfbox{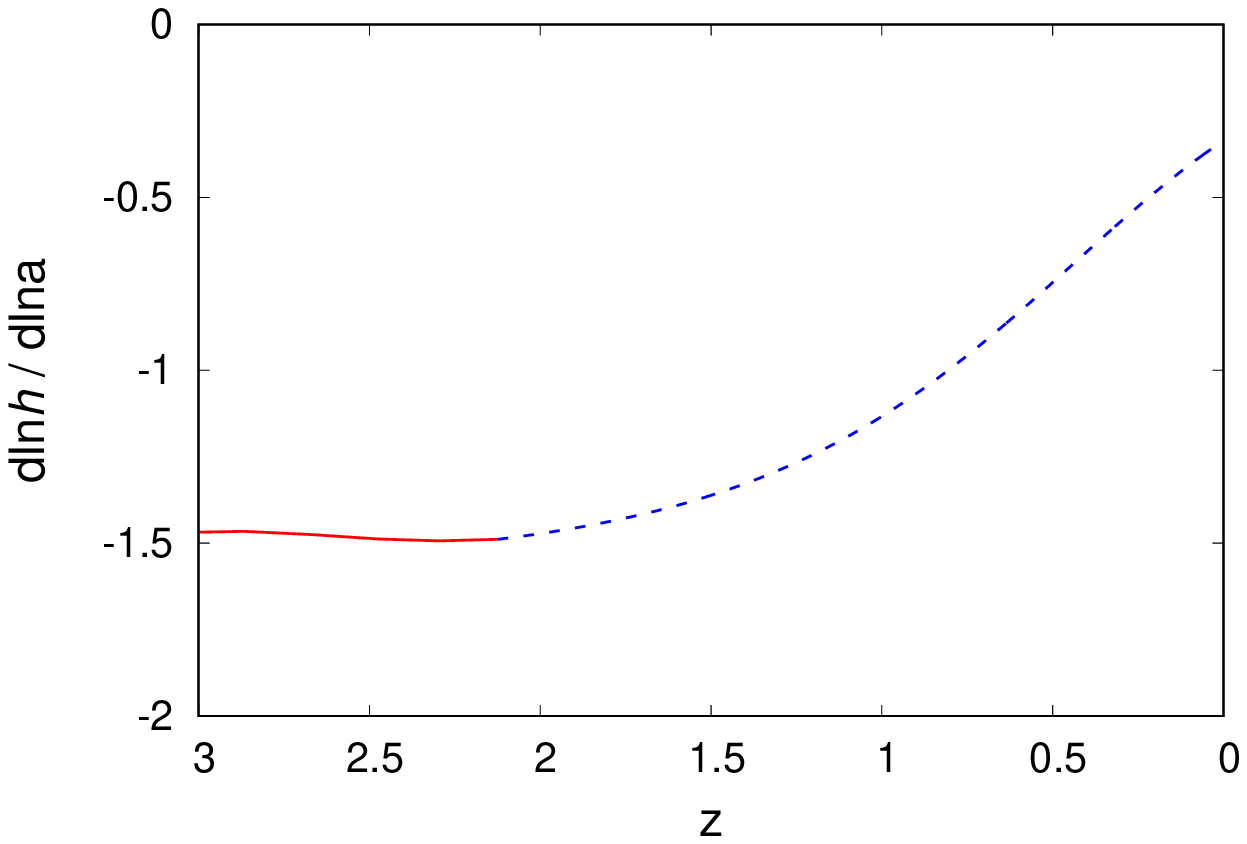}}
\epsfxsize=8. cm \epsfysize=6 cm {\epsfbox{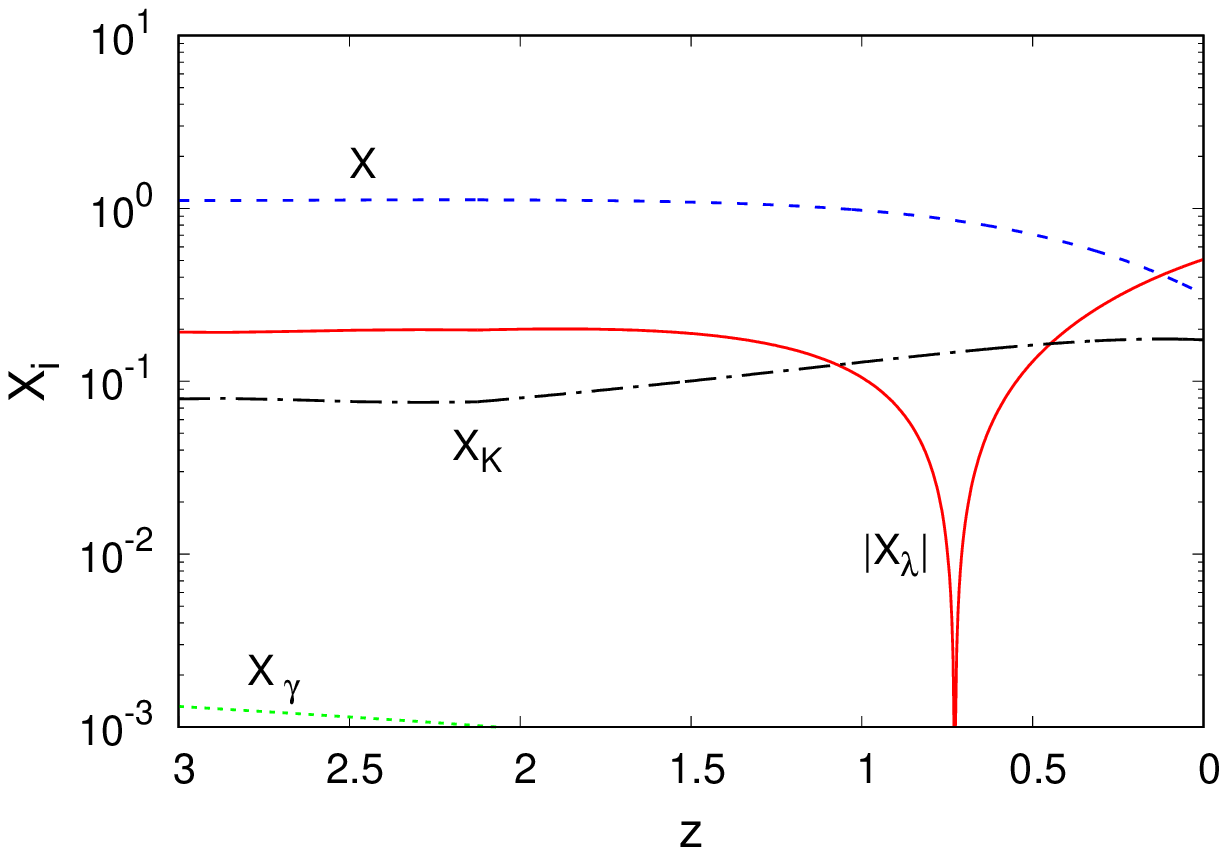}}
\end{center}
\caption{
{\it Upper panel:} derivative $d\ln \hh/d\ln a$ during the late matter era
and the dark energy era.
{\it Lower panel:} contributions $X_i$ to the Friedmann equation.
}
\label{fig_dlnh-Xi-DE}
\end{figure}

For the kinetic and coupling functions (\ref{eq:K-DE})-(\ref{eq:A-DE}),
the equations of motion (\ref{eq:dS-dphi-hat})-(\ref{eq:dS-dlambda-hat}) become
\beq
\frac{d^2\lambdabbar}{d\eta^2} = - \left( 3+\frac{d\ln \hh}{d\eta} \right)
\frac{d\lambdabbar}{d\eta}
\label{eq:dS-dphi-DE}
\eeq
and
\beqa
\frac{d^2\varphi}{d\eta^2} & = & \frac{A^4}{K_Y \hh^2}
- \left( 3+\frac{d\ln\hh}{d\eta} \right) \frac{d\varphi}{d\eta} \nonumber \\
&& - \frac{6 (1-\gamma) \gamma K_X}{K_Y} X^{\gamma-1} \frac{d\lambdabbar}{d\eta} .
\label{eq:dS-dlambda-DE}
\eeqa
The linear equation (\ref{eq:dS-dphi-DE}) shows a constant mode and a decaying mode,
with $d\lambdabbar/d\eta \propto e^{-3\eta}/\hh$. It fully determines $\lambdabbar$,
which is no longer coupled to $\varphi$.
The scalar field $\varphi$ is governed by Eq.(\ref{eq:dS-dlambda-DE}).
If the last term is positive or negligible, $\varphi$ will keep growing with time and
the Universe always remains in the accelerated expansion phase, unless the kinetic
and coupling functions again change form at higher values of $\varphi$.
On the other hand, if the last term is sufficiently large and negative, $\varphi$
may decrease in the future and finally leave the regime (\ref{eq:K-DE})-(\ref{eq:A-DE}),
to enter again the matter-era regime.
We do not investigate these various possibilities as they depend on the form of the
kinetic and coupling functions for arguments that cannot be probed by observations
(at least at the background level).

We show in Figs.~\ref{fig_lambda-phi-DE} and \ref{fig_dlnh-Xi-DE} our numerical results,
with the transition time $a_{\rm DE} = 0.3$.
In this simple example, we can see that $\varphi$ turns around and is decreasing
at $z=0$, so that the dark energy era would not last forever.
The derivative of the Hubble expansion rate, $d\ln\hh /d\eta$, grows from the
matter-era value $3/2$ towards zero, associated with a cosmological constant era.
The contribution from the matter component to the Friedmann equation decreases
from about unity to $0.32$, which corresponds to the value of the cosmological
parameter $\Omega_0$ today.

In this paper, we do not try to match the expansion history shown in
Fig.~\ref{fig_dlnh-Xi-DE} to observational data. To do so one would need to implement
a smooth transition for the kinetic and coupling function, tuned so as to reproduce
the observed Hubble diagram. Better still, one should first make the scalar field
negligible during the late matter era, below the contribution
(\ref{eq:Friedmann-matter-era-scalar}) associated with our power-law kinetic
function. This would allow one to naturally implement the transition to a constant
$\lambda$ at much earlier redshifts, so that at low $z$ the dynamics becomes
identical to the standard $\Lambda$CDM cosmology.
Of course, a much earlier transition can also be achieved without changing
(\ref{eq:Friedmann-matter-era-scalar}), by tuning the transition such that the
constant value $\lambdabbar$ reached at late times is much smaller than the one
achieved at the beginning of the transition. However, this requires some amount
of tuning, in proportion to the ratio between the initial and final values of
$\lambdabbar$.
We leave a detailed study of these points for future works.

Scalar-tensor theories  often give rise to long-range  fifth-forces, which are strongly constrained by solar system data \cite{Bertotti:2003rm}. In our model no long range force is present at low redshift as the coupling to matter $\frac{dA}{d\varphi}$ vanishes identically. Hence no local deviation from General Relativity appears.

\section{Inflation era}
\label{sec:Inflation-era}

\subsection{Accelerated expansion stage}
\label{sec:Accelerated-expansion}

In our numerical computation of the radiation era, in section~\ref{sec:radiation-numerical-1},
we started at early times $a \lesssim 10^{-25}$ close to the solution (\ref{eq:scalar-rad}).
We did not specify how this initial condition is achieved.
This can be considered as beyond the scope of our model, if we consider that it is
a low-energy effective Lagrangian that only applies after the inflation era.
However, it is interesting to see how the inflationary era could also be incorporated
within our framework. Because it corresponds to an accelerated expansion,
driven by an effective cosmological constant that is usually associated with the
value of the inflation potential during its slow-rolling phase, the cancellation mechanism
of the vacuum energy density described in section~\ref{sec:cancellation-radiation} must not
apply to this epoch, or be inefficient.
As for the dark-energy era discussed in section~\ref{sec:end-cancellation},
this naturally happens when $dA/d\varphi$ is zero or negligible, so that the equations of motion
(\ref{eq:dS-dphi-hat})-(\ref{eq:dS-dlambda-hat}) only depend on derivatives of $\lambda$.
Then, generically there is a constant mode for $\lambda$, which no longer systematically
runs towards $\Omega_{\rm vac}$ and compensates the vacuum energy density in the Friedmann
equations.

Thus, let us consider the case of a standard kinetic function of the form
\beq
K(\varphi;X,Y,Z) = K_X  X + K_Y Y  ,
\label{eq:K-Inf-1}
\eeq
and constant coupling function
\beq
\varphi < \varphi_{\rm I} : \;\;\; A(\varphi) = A_{\rm I}  .
\label{eq:A-Inf-1}
\eeq
The kinetic function (\ref{eq:K-Inf-1}) has a standard form, in the sense that it is a quadratic
polynomial in $\partial\lambda$ and $\partial\varphi$. For simplicity, we put the term $K_Z$
to zero. In fact, because the system converges to $d\lambda/d\eta=0$, the term
$K_X X$ plays no role and $K_X$ can take any value, including zero
(it disappears from the scalar field equations of motion, and it gives
a vanishing contribution to the Friedmann equations for $d\lambda/d\eta=0$).

We also consider an alternative scenario to the standard inflaton model, where the accelerated
expansion is due to the high-energy vacuum energy density $\Omega_{\rm vac}$,
and the end of the inflationary stage is due to a phase transition that decreases $\Omega_{\rm vac}$
while generating a nonzero radiation component, in a manner similar to the EW and QCD phase
transitions described in section~\ref{sec:phase}.
Then, one would need to ascribe the small metric fluctuations that give rise to the CMB anisotropies
and large-scale structures to other spectator fields, which do not drive the background
expansion \cite{Mazumdar:2010sa}.
Here we do not study these points in details, which go beyond the scope of this paper, and only sketch
how an inflationary era could be connected to the later radiation era.

With the kinetic and coupling functions (\ref{eq:K-Inf-1})-(\ref{eq:A-Inf-1}), the equations of motion
(\ref{eq:dS-dphi-hat})-(\ref{eq:dS-dlambda-hat}) give
\beq
\frac{d^2\lambdabbar}{d\eta^2} + \left( 3+\frac{d\ln \hh}{d\eta} \right)
\frac{d\lambdabbar}{d\eta} = 0 ,
\label{eq:dS-dphi-Inf-1}
\eeq
\beq
\frac{d^2\varphi}{d\eta^2} + \left( 3+\frac{d\ln\hh}{d\eta} \right) \frac{d\varphi}{d\eta}
=  \frac{A^4_{\rm I}}{K_Y \hh^2} ,
\label{eq:dS-dlambda-Inf-1}
\eeq
while the Friedmann equation (\ref{eq:EE00-hat}) reads as
\beq
\hh^2 = A^4 ( \Omega_{\rm vac 0} - \lambda) + \frac{\Omega_{\gamma 0}}{a^4}
+ K_X X + K_Y Y ,
\hspace{1cm}
\label{eq:EE00-Inf-1}
\eeq
where we set the energy density of nonrelativistic matter to zero.
We also set the initial radiation density to zero and the vacuum energy density to
a constant value $\Omega_{\rm vac I}$,
\beq
\Omega_{\gamma \rm I} = 0 , \;\;\; \Omega_{\rm vac} = \Omega_{\rm vac I} ,
\eeq
Then, we have the constant-$\lambda$ solution
\beq
\lambda = \lambda_{\rm I} , \;\;\; \lambdabbar = \lambda_{\rm I} - \Omega_{\rm vac I} , \;\;\;
\varphi = \varphi_I + \frac{A_{\rm I}^4}{3K_Y \hh_{\rm I}^2} ( \eta-\eta_{\rm I} ) ,
\label{eq:lambda-I-phi-I}
\eeq
where we assumed that the decaying modes $\propto e^{-3\eta}$ of $\lambda$ and $\varphi$
have had time to become negligible, and the Hubble expansion rate is
\beq
\hh^2_{\rm I} = - A_{\rm I}^4 \lambdabbar_{\rm I} , \;\;\; \mbox{with} \;\;\; X=0 , \;\; Y= 0 .
\eeq
Then, we assume that the constant values $\Omega_{\rm vac I}$ and $\lambda_{\rm I}$
are such that $\lambdabbar_{\rm I}$ is negative and $\hh_{\rm I}$ is of the order of the expected
inflationary scale.
We also take $K_Y>0$, so that $\varphi$ grows with time.
Indeed, we wish $\varphi$ to play the role of a clock, which triggers different cosmic regimes
through the dependence of $A$ and $K$ on $\varphi$. Since $\varphi$ is mostly growing
during the radiation, matter and dark-energy eras, it is convenient to have $\varphi$
growing during the inflationary era to avoid multivalued functions.
As noticed above, this solution does not depend on $K_X$, which can take any value.

\begin{figure}
\begin{center}
\epsfxsize=8. cm \epsfysize=6 cm {\epsfbox{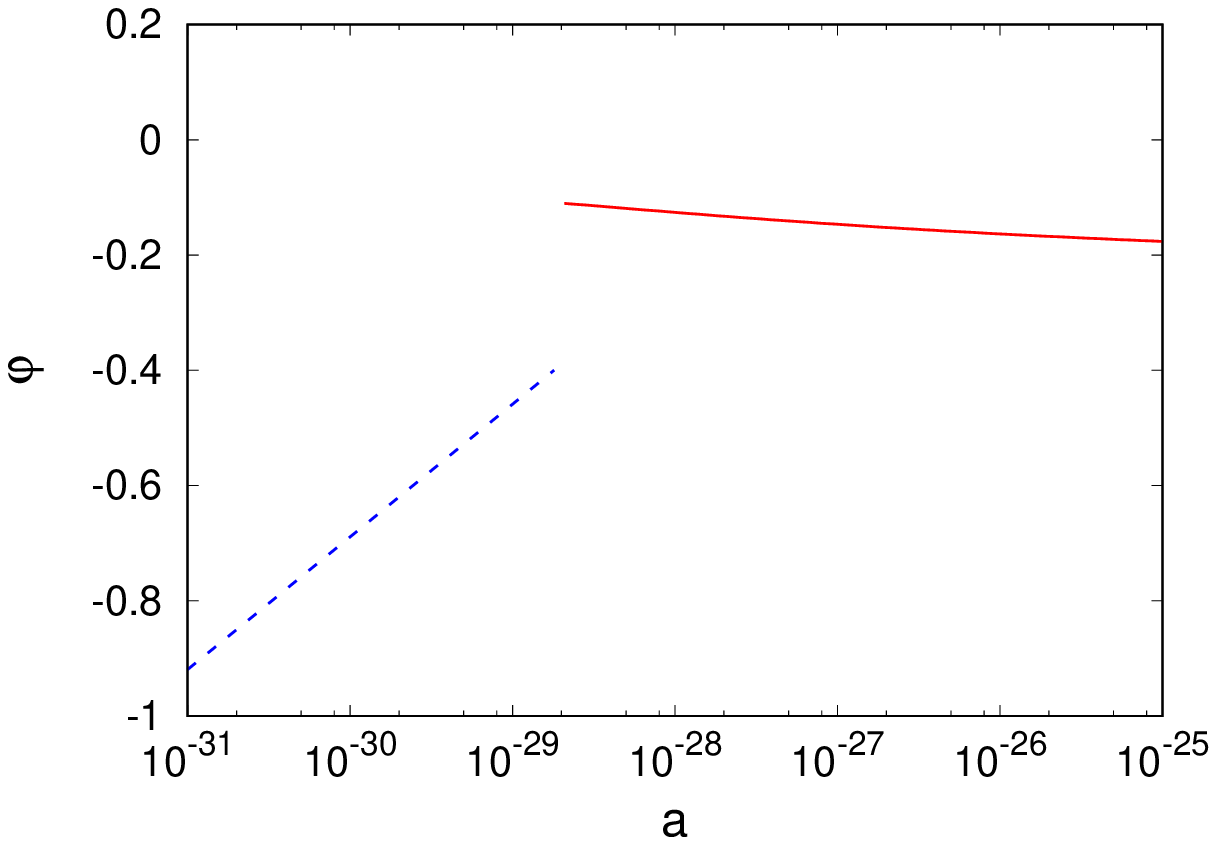}}
\epsfxsize=8. cm \epsfysize=6 cm {\epsfbox{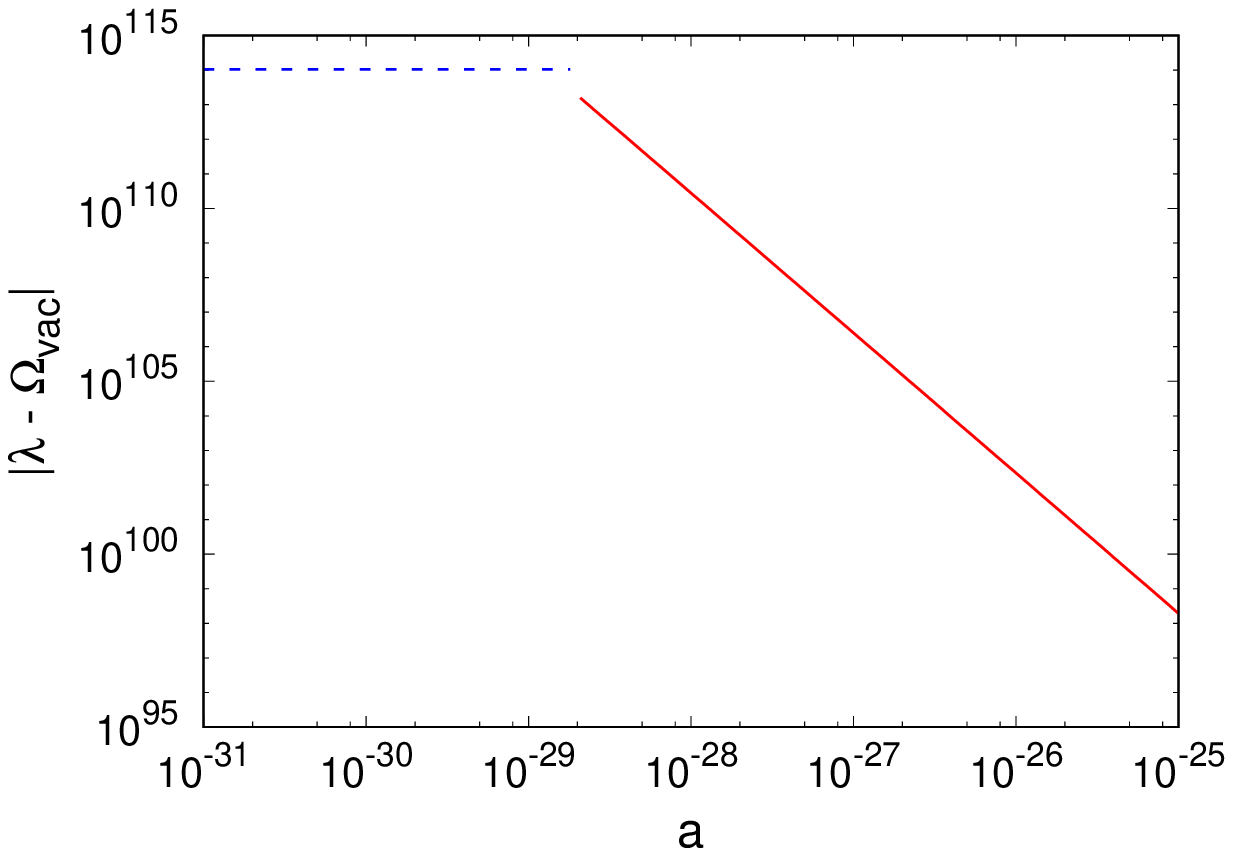}}
\end{center}
\caption{
{\it Upper panel:} scalar field $\varphi$ during the late inflation era (blue dashed line)
and the early radiation era (red solid line).
{\it Lower panel:} difference $\lambdabbar=\lambda-\Omega_{\rm vac0}$.
}
\label{fig_lambda-phi-Inf}
\end{figure}

\begin{figure}
\begin{center}
\epsfxsize=8. cm \epsfysize=6 cm {\epsfbox{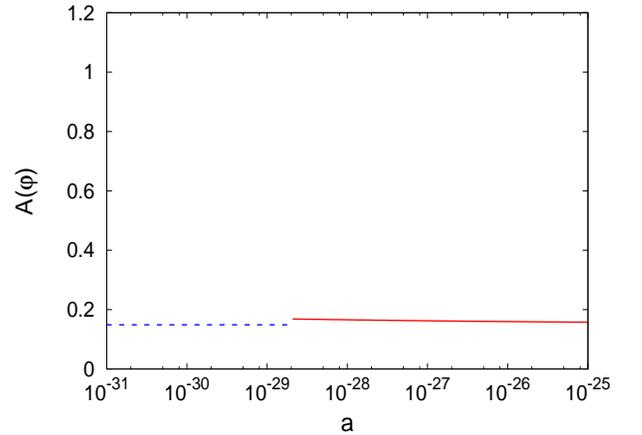}}
\end{center}
\caption{
Conformal factor $A(\varphi)$ during the late inflation era (blue dashed line)
and the early radiation era (red solid line).
}
\label{fig_A-Inf}
\end{figure}

\begin{figure}
\begin{center}
\epsfxsize=8. cm \epsfysize=6 cm {\epsfbox{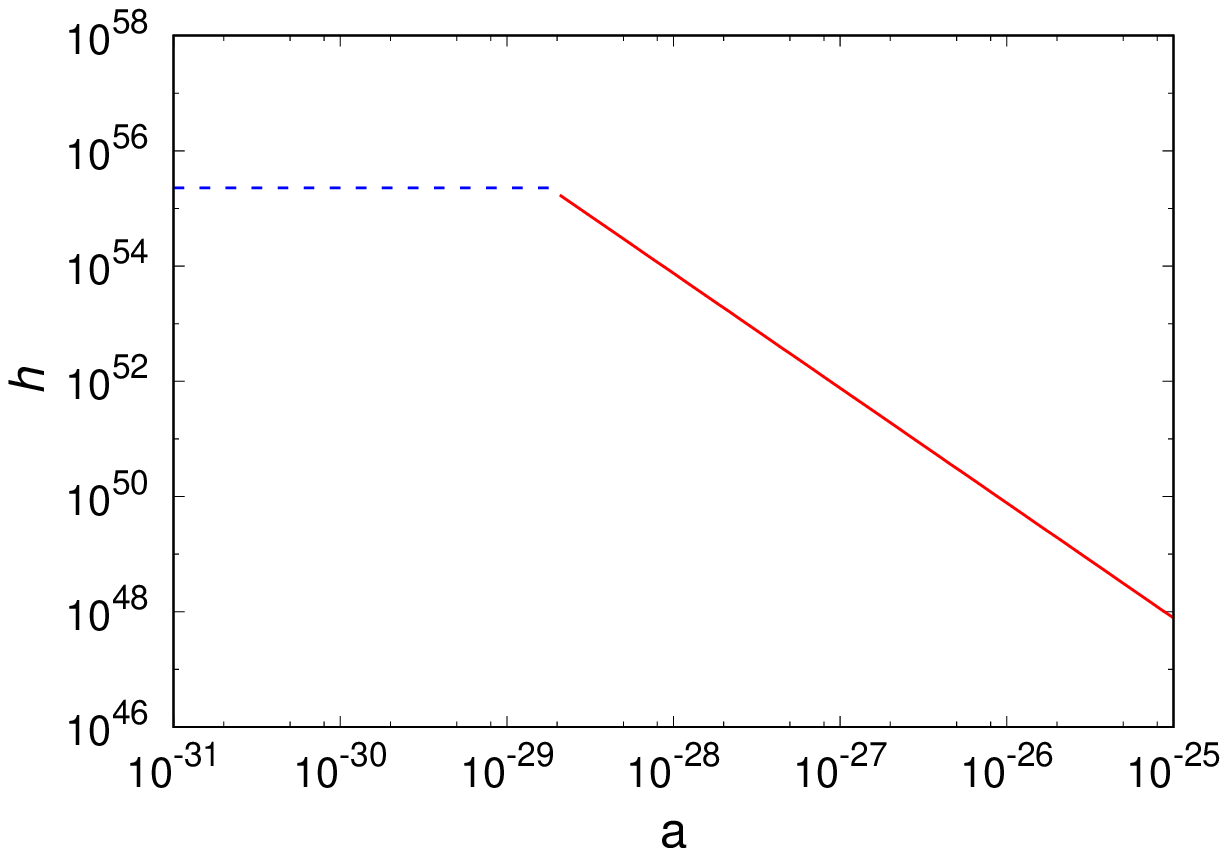}}
\epsfxsize=8. cm \epsfysize=6 cm {\epsfbox{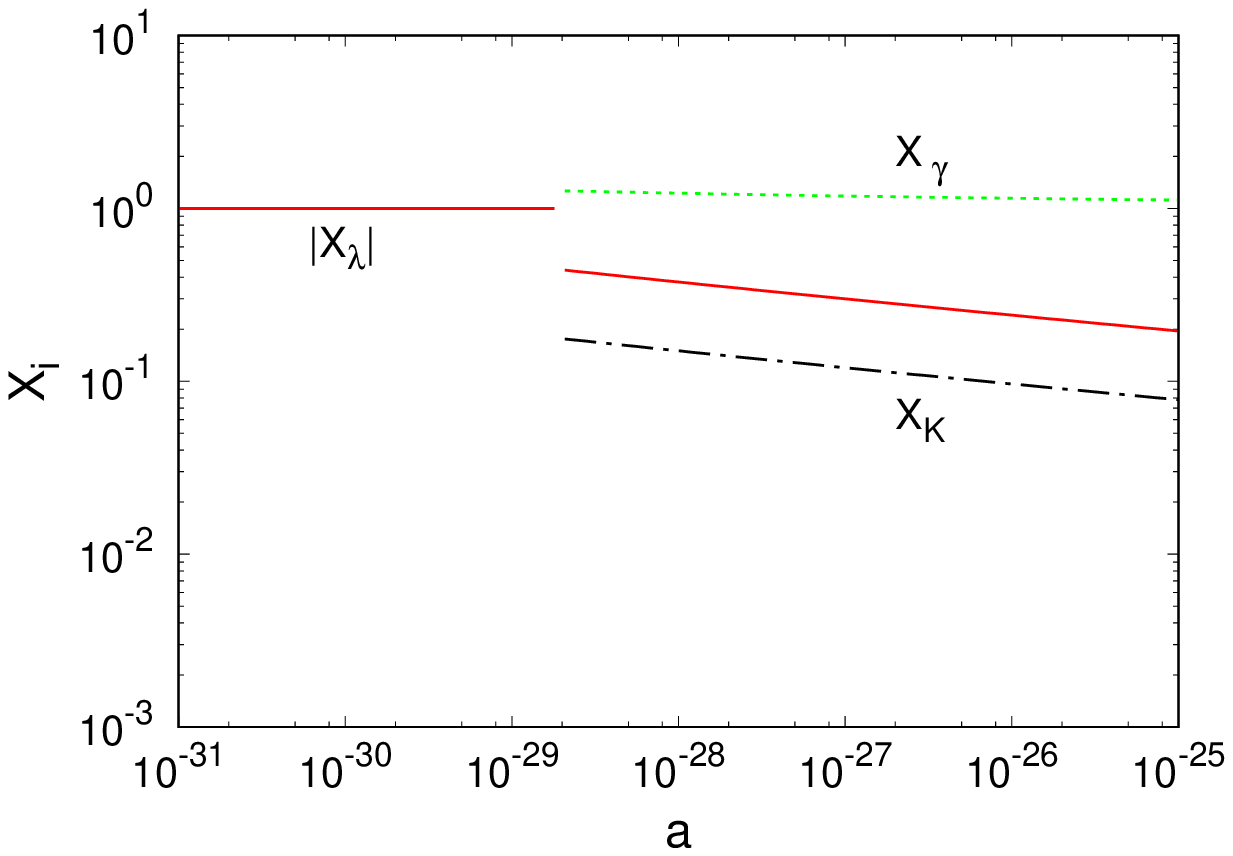}}
\end{center}
\caption{
{\it Upper panel:} reduced Hubble expansion rate $\hh$ during the late inflation era (blue dashed line)
and the early radiation era (red solid line).
{\it Lower panel:} contributions $X_i$ to the Friedmann equation.
}
\label{fig_h-Xi-Inf}
\end{figure}

\subsection{End of the accelerated expansion stage}
\label{sec:End-of-the-accelerated-expansion}

We assume that the inflationary epoch ends at the time $\eta_{\rm I}$ through a phase transition,
which suddenly decreases $\Omega_{\rm vac}$ while increasing the radiation component
$\Omega_{\gamma 0}$. As for the matter phase transitions studied in section~\ref{sec:phase},
we consider a simplified treatment where this transition is homogeneous and instantaneous.
We also assume that the kinetic and coupling functions show a transition at the same time
to the radiation-era forms (\ref{eq:K-rad}) and (\ref{eq:A-rad-def}), with the coefficients
$A_{\star},\nu_A,K_X,\nu_X$ and $\gamma$ that we used in section~\ref{sec:radiation-numerical-1}
for our numerical computation of the radiation era.
In particular, as $\nu_A$ is no longer zero the scalar field $\lambda$ will not remain constant
but decay as in Fig.~\ref {fig_lambda-phi-rad}.
We can imagine a scenario where these two events are related,
associated with $\varphi$ reaching the critical value $\varphi_{\rm I}$,
or discard the change of the vacuum energy density and only relate the end
of the inflationary stage to the change of the kinetic and coupling functions.
The term $\frac{\partial K}{\partial Y} \frac{d\lambda}{d\eta}$ in Eq.(\ref{eq:dS-dphi-hat})
is continuous as it is zero on both sides of the transition.
The continuity of the term $\frac{\partial K}{\partial X} \frac{d\lambda}{d\eta}
+ \frac{\partial K}{\partial Y} \frac{d\varphi}{d\eta}$ in Eq.(\ref{eq:dS-dlambda-hat}) gives the
junction condition
\beq
K_{Y_1} \left. \frac{d\varphi}{d\eta} \right|_1 = K_{X_2} e^{\nu_{X_2}\varphi_2} \gamma_2
X_2^{\gamma_2-1} \left. \frac{d\lambdabbar}{d\eta} \right|_2 .
\eeq
The scalar field $\lambda$ and the Hubble expansion rate are continuous at the transition.
Because the vacuum energy density drops at the transition by a quantity
\beq
\Delta \Omega_{\rm vac} = - \alpha_{\rm I} \frac{\hh_{\rm I}^2}{A_{\rm I}^4} , \;\;\;
\alpha_{\rm I} > 0 ,
\eeq
the difference $\lambdabbar$ grows by $-\Delta\Omega_{\rm vac}$, hence,
\beq
\lambdabbar_2 = (\alpha_{\rm I}-1) \frac{\hh_{\rm I}^2}{A_{\rm I}^4} .
\label{eq:lambdabbar-2-Inf-2}
\eeq
The radiation density after the transition is then given by
\beq
\Omega_{\gamma 2} = a_{\rm I}^4 \left[ \hh_{\rm I}^2 + \frac{A_2^4}{A_{\rm I}^4} (\alpha_{\rm I} -1)
h_{\rm I}^2 - (2\gamma_2-1)K_2 \right] .
\eeq
As for the matter phase transitions (\ref{eq:Omegarad-pt}), the drop of the vacuum energy density
is transferred to both radiation and scalar field components, because of the discontinuous
scalar field couplings.

\subsection{Numerical computation}
\label{sec:numerical-inflation}

We show in Figs.~\ref{fig_lambda-phi-Inf}-\ref{fig_h-Xi-Inf} a numerical computation of the scenario
described in the previous sections.
We take $H_{\rm I} = 10^{-5} M_{\rm Pl}$ for the Hubble expansion rate during the
inflationary era. This corresponds to $\hh_{\rm I} \simeq 10^{55}$.
At the transition, $\varphi$ and $\lambdabbar$ are discontinuous and next follow the evolution
that characterizes the radiation era analyzed in section~\ref{sec:Radiation-era}.
In practice, we choose the value of $A_{\rm I}$ so as to recover the numerical values
obtained in Fig.~\ref{fig_lambda-phi-rad} during the radiation era
(we can check on the figures that they match at $a \sim 10^{-25}$ where the plots
overlap).
The value of the kinetic function coefficient $K_Y$ is irrelevant, as it only determines
the value of $d\varphi/d\eta$ at early times.
Indeed, only the combination $K_Y \frac{d\varphi}{d\eta}$ enters the equations; therefore,
$K_Y$ and $d\varphi/d\eta$ are degenerate.
Because $\varphi$ slightly decreases during the early radiation era, as seen in
Fig.~\ref{fig_lambda-phi-rad}, and we want $A(\varphi)$ to be single-valued,
we take $\varphi$ discontinuous at $\eta_{\rm I}$ so that it is safely below radiation-era values
during the full inflationary stage.
The value reached just before $\eta_{\rm I}$ is a free parameter, and we could also make
$\varphi$ continuous by changing the slope during the radiation era to a small positive value.
The conformal factor $A(\varphi)$ also shows a small discontinuity at $\eta_{\rm I}$,
as seen in Fig.~\ref{fig_A-Inf}.
Whereas $\lambdabbar$ is negative before $\eta_{\rm I}$, it is positive after the transition
thanks to the drop of the vacuum energy density, with $\alpha_{\rm I}>1$ in
Eq.(\ref{eq:lambdabbar-2-Inf-2}). Our numerical results correspond to $\alpha_{\rm I} \simeq 1.3$.
As displayed in Fig.~\ref{fig_h-Xi-Inf}, the Hubble expansion rate is constant during the inflationary
stage and next decreases almost as $a^{-2}$. The radiation component is dominant
after $\eta_{\rm I}$ while the scalar field contributions to the Friedmann equation are subdominant
and decrease slightly faster than the radiation component, as described in section~\ref{sec:Radiation-era}.

As explained above, this numerical computation is only meant as an example for a transition
from the inflationary to the radiation era. It does not address the beginning of the inflationary
era itself. Also, the transition to the radiation era would deserve more detailed studies.
This is only one of the possible scenarios, and it should be possible to discard the change of
the vacuum energy density and to associate the transition to a change of the scalar field
coupling functions, which initiate the compensation mechanism described in
section~\ref{sec:cancellation-radiation}. Then, we would need to specify a reheating mechanism,
possibly associated with other fields or with the inflaton (which could replace the nonzero vacuum
energy density during the accelerated expansion phase), to generate the radiation component
that governs the subsequent radiation era.
A complete scenario must also provide the almost scale-invariant primordial fluctuations
that give rise to the CMB anisotropies and the formation of large-scale structures in the late
Universe, possibly through the quantum fluctuations of other spectator fields.
All these points go beyond the scope of this paper and are left to future works.
Alternatively, one can consider that the action (\ref{eq:S-def}) is only a low-energy effective
model, which does not apply to the inflationary era, or that the scalar fields
$\lambda$ and $\varphi$ play no role during the inflationary and early radiation eras
(by keeping $A$ constant throughout), so that the standard inflationary scenario applies
without any modification.

\section{Conclusion}
\label{sec:conclusion}

In this paper, we have presented a simple scenario that provides a cosmological
cancellation of the matter vacuum energy seen by gravity.
This ``conformal compensator model'' relies on a dynamical conformal rescaling
$A(\varphi)$ between
the Jordan-frame metric $\tilde{g}_{\mu\nu}$ seen by the matter Lagrangian
and the Einstein-frame metric $g_{\mu\nu}$ seen by gravity,
which is still given by the Einstein-Hilbert action of General Relativity.
When this factor $A$ is constant, we recover General Relativity, with
a possible nonzero value of the cosmological constant, associated with the
vacuum energy and a constant value of a second scalar field $\lambda$.
When the conformal factor $A$ has a nonzero first derivative and becomes time dependent,
it induces a coupling between $\lambda$ and the trace of the energy-momentum tensor,
such that $\lambda$ cancels
the vacuum energy density $\tilde{V}_{\rm vac}$, leading to a radiationlike
expansion of the Universe (in the Einstein frame).
This mechanism evades Weinberg's no-go theorem \cite{Weinberg:1988cp}
thanks to a time-dependent
background field $\varphi$ (even in the Minkowski limit).
This is natural in the cosmological setting. As the Universe is not static,
there is no reason to require static background fields.

This mechanism shares with the sequestering scenario \cite{Padilla_2014,Kaloper:2014dqa}
the key role
played by the conformal rescaling. However, the factor $A(\varphi)$
is no longer a global variable, set by the full history of the Universe,
but a dynamical field. This avoids causality problems, where the value of the
cosmological constant today depends on both the past and future histories
of the Universe. However, this makes the scenario more complex, as one needs
to follow the dynamics of these scalar fields.
On the other hand, making the cancellation mechanism dynamical offers several
advantages, as it allows us to link together the different eras of the expansion
history of the Universe.
Thus, both the early inflation and late dark-energy eras can be due to
periods where this cancellation mechanism is inefficient (e.g., the coupling
$dA/d\varphi$ is too weak), while the intermediary radiation and matter eras correspond
to periods where the mechanism is at play.
As the model naturally leads to cosmological expansions that are dominated
either by an effective cosmological constant or by a radiation component,
it could provide a first step to explain why the matter era is only a small temporary
period in the history of the Universe, as measured in the number of e-folds.

The explicit implementation presented in this paper is not complete nor fully satisfactory.
First, we only sketched a possible scenario for the inflationary stage and its
transition to the radiation era. Second, the matter era remains problematic.

If one insists on incorporating the inflation era within this framework,
much more detailed studies are needed
that address in particular the generation of the late-time radiation density,
e.g. through reheating mechanisms, and of the primordial almost scale-invariant
fluctuations that lead to the CMB anisotropies and large-scale structures,
e.g. through spectator fields.

Alternatively, one can introduce the usual inflaton field, responsible for both the
accelerated expansion and the primordial fluctuations.
In this minimal scenario, the scalar fields introduced in this paper play no role
and one simply recovers the standard cosmology. One only needs to make the cancellation
mechanism inefficient during this era. More precisely, $\lambda$ would screen
higher-energy vacuum energy densities before the inflationary stage but the mechanism
would stop during the inflationary stage, until later during the radiation era, before the
EW and QCD phase transitions.
This can be easily achieved by making $A(\varphi)$ constant during this period.
This possibility ensures that we recover the standard early-time cosmology.
The drawback is that the inflationary era and its end are no longer connected
to the scalar fields $\varphi$ and $\lambda$. They actually do not need to be,
but it would be elegant to connect closely the cancellation of the vacuum energy,
the inflationary era, and the dark energy era.

However, the most pressing issue is the treatment of the matter era.
The problem comes from the fact that the cancellation mechanism makes
the field $\lambda$ respond to both the vacuum energy density
$\tilde{V}_{\rm vac}$ and the nonrelativistic matter density $\tilde\rho$,
because it is coupled to the trace of the energy-momentum tensor.
This is not a problem in the sequestering models, because there $\lambda$ is a global
variable. It is coupled to the average over all spacetime of $T^\mu_\mu$,
which is dominated by the value at late times, set by the cosmological constant
or the final low-energy vacuum energy density, as matter and radiation components
are diluted away by the expansion of the Universe.
In other words, the vacuum energy density is distinguished as the constant component
that is left when all others have been diluted by the expansion.
In contrast, in the model presented in this paper, because the scalar field $\lambda$ is
dynamical, we would need to recognize the vacuum energy density on the spot,
at each moment in time. This is not possible, as there is always an ambiguity
(e.g., from an observational point of view) in the one-time splitting between a vacuum energy
component and a matter component (e.g., a slowly varying scalar field potential,
or a matter component with an intricate equation of state).
This is why in our model $\lambda$ gets coupled to $T^\mu_\mu$, which includes
both $-4\tilde{V}_{\rm vac}$ and $-\tilde\rho$, where $\tilde\rho$ is the density of
nonrelativistic matter.

Within the framework defined by the simple kinetic functions (\ref{eq:K-rad}),
we have seen that this problem can be circumvented in a natural manner by making
the field $\lambda$ track the matter density, with an amplitude smaller than unity
so that it is not the dominant component.
In this manner, the coupling to matter actually helps to make sure that the field
$\lambda$ does not converge to a constant, much above the matter density,
which would lead to a dark energy era immediately following the radiation era,
without any matter era.

However, the explicit example computed in this paper suffers from two shortcomings.
First, as stability requirements are not identical in the radiation and matter eras,
the kinetic function $K$ must change form somewhat before the onset of the matter era,
i.e. the exponent $\nu_X$ in Eq.(\ref{eq:K-rad}) must decrease from about $1/3$
to about $-0.4$. This must occur after the latest matter phase transition
($a \sim 10^{-12}$ at $T_{\rm QCD}$) and before the radiation-matter equality
($a \sim 10^{-4}$). This implies a mild coincidence problem.
Moreover, we found that numerically the transition from the radiation to the matter
era is quite delicate to ensure the convergence to the late-time tracking solution.
Because the equations of motion are nonlinear, several branches of solutions can exist,
which can also lead to strong instabilities, and one does not always end up in the
branch that is similar to the standard cosmology.
Second, and more importantly, the contribution from the scalar fields to the Friedmann
equation is about $14\%$ in the radiation era, with our tracking solution.
This is probably too large to be consistent with observations, although a detailed
study would be needed to take into account degeneracies once we go beyond
the standard $\Lambda$CDM cosmology.
Thus, it would be desirable to obtain solutions that can simultaneously recover
the radiation and matter eras and that give a scalar-field contribution to the Hubble expansion
rate that is at the percent level or below.
This would in turn lead to a dark-energy era that closely mimics a cosmological constant,
the field $\lambda$ having converged to a constant at much earlier times.

We can hope that more complex Lagrangians (and maybe additional fields)
could solve these problems.
We leave such investigations for future works.

\acknowledgements
We thank E.V. Linder and A. Padilla for comments on the draft of this paper.

\appendix

\section{Numerical implementation at the end of the radiation era}
\label{app:jumps}

In practice, to implement the variation of the parameters of the kinetic function (\ref{eq:K-rad}),
instead of using a smooth function that interpolates between the different regimes, we introduce
small jumps of $\sigma$ at several time steps (i.e., we discretize the transition).
This allows us to keep the simple equations of motion (\ref{eq:dS-dphi-rad})-(\ref{eq:dS-dlambda-rad})
between these discrete events.
However, we need to make sure that the equations of motion remain satisfied across these discontinuities.
The constraint equation (\ref{eq:dS-dphi-hat}) (with $K_Y=K_Z=0$) does not give any junction condition,
as there is no kinetic term over $\varphi$, which instantaneously responds to changes of other quantities.
In contrast, the equation of motion (\ref{eq:dS-dlambda-hat}) implies that we must keep
$\frac{\partial K}{\partial X} \frac{d\lambdabbar}{d\eta}$ continuous across the
boundaries.

For the very small jumps before and after $a \simeq 10^{-5}$, which model a continuous decrease
of $\sigma$, we proceed as follows. We keep $\varphi$ and $\lambdabbar$ continuous.
Then, Eq.(\ref{eq:dS-dphi-rad}) implies that its left-hand side is continuous.
Together with the continuity of $\frac{\partial K}{\partial X} \frac{d\lambdabbar}{d\eta}$,
this gives the junction condition
\beq
\left( \frac{d\lambdabbar}{d\eta} \right)_2 = \left( \frac{d\lambdabbar}{d\eta} \right)_1
\frac{\gamma_2 \nu_{X_1}}{\gamma_1 \nu_{X_2}} ,
\eeq
across small changes of the parameters $\gamma$ or $\nu_X$.
This also provides the change of the kinetic factor $X$, from $X_1$ to $X_2$.
Next, the continuity of the left-hand side of Eq.(\ref{eq:dS-dphi-rad})
leads to
\beq
K_{X_2} = K_{X_1} \frac{\nu_{X_1}}{\nu_{X_2}}
e^{(\nu_{X_1}-\nu_{X_2})\varphi} X_1^{\gamma_1} X_2^{-\gamma_2} .
\eeq
This provides a simple implementation of slowly-varying coefficients $\gamma$ or $\nu_X$,
while satisfying at all times the equations of motion of the scalar fields.
As we are still inside the radiation era, we neglect the effects of the small jumps of $\hh$,
because the Hubble expansion rate is mostly determined by the radiation density.

For the finite jump that occurs at $a \simeq10^{-5}$, we proceed in a different manner.
We keep $\lambdabbar$ continuous as well as $\frac{\partial K}{\partial X} \frac{d\lambdabbar}{d\eta}$,
to satisfy the equation of motion across the discontinuity.
However, we allow $\varphi$ to be discontinuous to make sure that we ``shoot'' into the basin of
attraction of the solution (\ref{eq:scalar-matter}).

\bibliography{ref1}   

\end{document}